\documentclass[11pt,a4paper]{article}
\pdfoutput=1

\usepackage{jcappub}
\usepackage{natbib}

\title{Elliptical galaxies kinematics within general relativity with renormalization group effects}

\author[a]{Davi C. Rodrigues}

\affiliation[a]{Departamento de F\'isica, CCE, 
Universidade Federal do Esp\'irito Santo, 29075-910, Vit\'oria, ES, Brazil }

\emailAdd{davirodrigues.ufes@gmail.com}

\abstract{The renormalization group framework can be applied to Quantum Field Theory on curved space-time, but there is no proof whether the beta-function of the gravitational coupling indeed goes to zero in the far infrared or not. In a recent paper \cite{Rodrigues:2009vf} we have shown that the amount of dark matter inside spiral galaxies may be negligible if a small running of the General Relativity coupling $G$ is present ($\delta G/G_0 \lesssim 10^{-7}$ across a galaxy). Here we extend the proposed model to elliptical galaxies and present a detailed analysis on the modeling of  NGC 4494 (an ordinary elliptical) and NGC 4374 (a giant elliptical). In order to compare our results to a well known alternative model to the standard dark matter picture, we also evaluate  NGC 4374 with MOND. In this galaxy MOND leads to a significative discrepancy with the observed velocity dispersion curve and has a significative tendency towards tangential anisotropy. On the other hand, the  approach based on the renormalization group and general relativity (RGGR) could be applied with good results to these elliptical galaxies and is compatible with lower mass-to-light ratios (of about the Kroupa IMF type).}


\begin{document}

\maketitle

\renewcommand{\arraystretch}{1.3}
\renewcommand{\vec}[1]{{\bf #1}}
\renewcommand{\Re}{\,\mbox{Re}\,}
\renewcommand{\Im}{\,\mbox{Im}\,}

\def\<{\left \langle}
\def\>{\right \rangle}
\def\[{\left\lbrack}
\def\]{\right\rbrack}
\def\({\left(}
\def\){\right)}
\newcommand{\be}{\begin{equation}}
\newcommand{\ee}{\end{equation}}
\newcommand{\ea}{\end{eqnarray}}
\newcommand{\ba}{\begin{eqnarray}}
\newcommand{\prt}{{\partial}}
\newcommand{\diag}{\mbox{diag}}
\newcommand{\tr}{\mbox{tr}}
\newcommand{\grad}{\ensuremath{\vec{\nabla}}}
\newcommand{\bs}{\begin{sideways}}
\newcommand{\es}{\end{sideways}}
\newcommand{\chir}{{\chi^2_{\mbox{\tiny{red}}}}} 
\newcommand{\Newt}{N}
\newcommand{\Iso}{{\mbox{\tiny Iso}}}
\newcommand{\mond}{{\mbox{\tiny MOND}}}
\newcommand{\stvg}{{\mbox{\tiny STVG}}}
\newcommand{\IsoInf}{{\mbox{\tiny Iso} \infty}}
\newcommand{\dg}{\dagger} 
\newcommand{\ML}{$\Upsilon_*^V/ (\frac{M_\odot}{L_{\odot,\mbox{\tiny V}}})$}
\newcommand{\amond}{$\frac{a_0}{1.35\times10^{-8} \mbox{\tiny cm/s}^2}$}
\newcommand{\mnras}{Mon. Not. R. Astron. Soc.}
\newcommand{\aap}{Astronomy $\&$ Astrophysics}
\newcommand{\apjs}{ApJS}
\newcommand{\apj}{Astrophys. J.}
\newcommand{\aj}{Astron. J.}
\newcommand{\pasa}{PASA}
\newcommand{\nat}{Nature}
\newcommand{\RGGR}{{\mbox{\tiny RGGR}}}
\newcommand{\MOND}{{\mbox{\tiny MOND}}}
\newcommand{\Ser}{{\mbox{\tiny S}}}
\newcommand{\ext}{{\mbox{\tiny ext}}}

\def\beq{\begin{eqnarray}}
\def\eeq{\end{eqnarray}}
\def\ln{\,\mbox{ln}\,}
\def\Det{\,\mbox{Det}\,}
\def\det{\,\mbox{det}\,}
\def\tr{\,\mbox{tr}\,}
\def\diag{\,\mbox{diag}\,}
\def\Tr{\,\mbox{Tr}\,}
\def\sTr{\,\mbox{sTr}\,}
\def\Res{\,\mbox{Res}\,}

\def\lap{\Delta}
\def\sla{\!\!\!\slash}
\def\al{\alpha}
\def\bet{\beta}
\def\ch{\chi}
\def\ga{\gamma}
\def\de{\delta}
\def\vp{\varepsilon}
\def\ep{\epsilon}
\def\ze{\zeta}
\def\io{\iota}
\def\ka{\kappa}
\def\la{\lambda}
\def\na{\nabla}
\def\pa{\partial}
\def\ro{\varrho}
\def\rh{\rho}
\def\si{\sigma}
\def\om{\omega}
\def\ph{\varphi}
\def\ta{\tau}
\def\th{\theta}
\def\te{\vartheta}
\def\up{\upsilon}
\def\Ga{\Gamma}
\def\De{\Delta}
\def\La{\Lambda}
\def\Si{\Sigma}
\def\Om{\Omega}
\def\Te{\Theta}
\def\Th{\Theta}
\def\Up{\Upsilon}

\section{Introduction} \label{intro}

Currently there is a large body of data coming from cosmological and astrophysical observations that is mostly consistent with the existence of dark matter.  Such observations also suggest that the hypothesized particles that constitute dark matter have very small cross section and typically  travel  much slower than light. These lead to the cold dark matter (CDM) framework, which is one of the pillars of the current standard cosmological model $\Lambda$CDM.

It is not only tempting, but mandatory to check if such dark matter particles exist (by detecting them in laboratory based experiments, for instance) and also to check if the gravitational effects that lead to the dark matter hypothesis could follow from a more detailed and complete approach to gravity. The effects of pure classical General Relativity at galaxies have been studied for a long time and, considering galaxy kinematics, the differences between General Relativity and Newtonian gravity are negligible (for a dispute on the latter, see however Ref. \cite{Cooperstock:2006dt}).

One of the subjects which attract significant attention currently is  modifications of General Relativity in order to replace either dark matter or dark energy. When all the astrophysical and cosmological data are considered, there is currently no particular model in this class that was proved to be as successful as the $\Lambda$CDM model, however diverse nontrivial achievements were accomplished by such new approaches to gravity; whilst the $\Lambda$CDM model has its own difficulties and problems, particularly at the galactic scale\cite{Gentile:2004tb,  2009NJPh...11j5029P,  2010Natur.465..565P,  Kroupa:2012qj}. Probably the most well known success of the modified gravity approach relies on galaxy kinematics, where MOND\cite{1983ApJ...270..365M,1983ApJ...270..371M,Milgrom:2003ui} is largely the most cited example and achieved success in some areas where the $\Lambda$CDM results are at least unclear \cite{2012AJ....143...40M, 2007A&A...472L..25G}. There are examples in the cosmological realm as well. In particular, considering the evolution of linear perturbations and its comparison to the CMB and the LSS data, results either close or identical to $\Lambda$CDM results can be found in different scenarios  \cite{Banados:2008fj, Batista:2011nu, Moffat:2009cv, Skordis:2005xk}. The Bullet Cluster \cite{Clowe:2006eq} was some times cited as a definitive proof on the existence of dark matter, however this system can also be modeled from a modified gravity perspective \cite{Brownstein:2007sr, Dai:2008sf}. These phenomenological results motivate the use of gravitational theories that are different from General Relativity in its standard form.
  
Independent on whether gravity should be or should not be quantized, we know that the matter fields should. It is well known that the renormalization group 
 can be extended to quantum field theory (QFT) on curved space time (e.g., Refs. \cite{Buchbinder:1992rb,Shapiro:2008sf,Shapiro:2009dh}). In particular, concerning 
the high energy (UV) behavior, there is hope that the running of $G$ may converge to a non-Gaussian fixed point  in accordance with the asymptotic  safety approach \cite{Niedermaier:2006wt, Weinberg:2009wa}. Our present concern is, however, not about the UV completeness, but with the behavior of $G$ in the far infrared  (IR) regime. 

In the far IR regime of quantum electrodynamics, one finds classical electromagnetism, and hence no renormalization group running of its coupling constant. This behavior is in accordance with the Appelquist-Carazzone decoupling \cite{Appelquist:1974tg} (see Ref.\cite{Goncalves:2009sk} for a recent derivation). In the case of gravity (in the context of QFT in curved space time) the same effect of decoupling has been obtained for the higher derivative terms in the gravitational action \cite{Gorbar:2002pw,Gorbar:2003yt}. Currently, it remains unclear whether the Einstein-Hilbert action coupling parameter  behaves as a constant at the far IR or not. Since in pure theoretical grounds with no additional hypothesis it is  hard to advance in this direction, this possibility of General Relativity deviation has been developed on different grounds a number of times before, e.g. \cite{Goldman:1992qs, Bertolami:1993mh, Dalvit:1994gf, Bertolami:1995rt,   Shapiro:2004ch,  Reuter:2004nx, Reuter:2007de}.

In  \cite{Rodrigues:2009vf} we presented new results on the application of renormalization group corrections to General Relativity in the astrophysical domain. Previous attempts to apply this picture to galaxies have considered for simplicity point-like galaxies (e.g., \cite{Shapiro:2004ch, Reuter:2004nx}). We extended previous considerations by identifying a proper renormalization group energy scale $\mu$  and by evaluating the consequences considering the observational data of disk galaxies. We proposed the existence of a relation between $\mu$ and the local value of the Newtonian  potential (this relation was reinforced  afterwards from a different approach, by using a scale setting formalism \cite{Domazet:2010bk}). With this choice, the renormalization group-based approach (RGGR) was capable to mimic dark matter effects with great precision. Also, it is remarkable that this picture induces a very small variation on the gravitational coupling parameter $G$, namely a variation of about $10^{-7}$ of its value across a galaxy (depending on the matter distribution). We call our model RGGR, in reference to renormalization group effects in General Relativity.

The main purpose of this work is to extend the analysis of \cite{Rodrigues:2009vf} (see also \cite{Shapiro:2004ch, Farina:2011me, Rodrigues:2011cq},\cite{Fabris:2012wg}) towards elliptical galaxies, testing the RGGR approach in this context. This extension is also of value to future RGGR applications, for instance in cluster of galaxies. Many of the elliptical galaxies behave as spherically symmetric stable systems which are mainly supported by velocity dispersions, in sharp contrast with the disk galaxies,  which are axially symmetric and are mainly supported by rotation velocity.  To this end, we deduce the additional effective mass introduced by RGGR, detail some aspects of its general behavior on elliptical galaxies and present a detailed fitting using recent observational data of the galaxies NGC 4374 (giant elliptical) and NGC 4494 (ordinary elliptical). Moreover we use exactly the same data of NGC 4374 to evaluate MOND and compare its results with the RGGR ones.\footnote{We do not do the MOND analysis for NGC 4494 since it was previously analyzed in Ref. \cite{Milgrom:2003ui} and this galaxy does not constitute a  hard test for MOND.} The numerical evaluation of the velocity dispersion (VD) curves and related procedures use a program made by us on Wolfram Mathematica.

\section{A brief review on RGGR}

The gravitational coupling parameter $G$ may  behave as a true constant in the far  IR limit, leading to standard General Relativity in such limit. Nevertheless, in the context of QFT in cuved space time, there is no proof on that. According to Refs. \cite{Shapiro:2004ch, Farina:2011me}, a certain logarithmic running of $G$  is a direct consequence of covariance and must hold in all loop orders.  Hence the situation is as follows: either there is no new gravitational effect induced by the renormalization group in the far infrared, or there are such deviations and the gravitational coupling runs as
\be
	 \beta_{G^{-1}} \equiv \mu \frac{dG^{-1}}{d \mu} = 2 \nu \,  \frac{M_{\mbox{\tiny Planck}}^{2}}{c \, \hbar} = 2 \nu G_0^{-1}.
	\label{betaG}
\ee
Equation (\ref{betaG}) leads to the logarithmically  varying $G(\mu)$ function,
\be
	\label{gmu}
	G(\mu) = \frac {G_0}{ 1 + \nu \ln(\mu^2/\mu_0^2)},
\ee
where $\mu_0$ is a reference scale introduced such that $G(\mu_0) =G_0 $. The constant $G_0$ is the gravitational constant as measured in the Solar System  (actually, there is no need to be very precise on where $G$ assumes the value of $G_0$, due to the smallness of the variation of $G$). The dimensionless constant $\nu$ is a phenomenological parameter which depends on the details of the quantum theory leading to eq. (\ref{gmu}). Since we have no means to compute the latter from first principles, its value should be fixed from observations. Even a small $\nu$ of about  $\sim 10^{-7}$ can lead to observational consequences at galactic scales. Note that the first possibility, namely of no new gravitational effects in the far infrared, corresponds to $\nu=0$.

The action for this model is simply the Einstein-Hilbert one in which $G$ appears inside the integral, namely,\footnote{We use the $(- + + +)$ space-time signature.}
\be
	S_{\mbox{\tiny RGGR}}[g] = \frac {c^3}{16 \pi }\int \frac {R  } G \, \sqrt{-g} \,  d^4x.
	\label{rggraction}
\ee
In the above, $G$ is an external scalar field, it satisfies (\ref{gmu}).  For a complete cosmological picture, $\Lambda$ is necessary and it also runs covariantly with the RG flow of $G$ \cite{Shapiro:2004ch, Reuter:2007de, Koch:2010nn}. In the above, the $\Lambda$ term was not written since its role is expected to be negligible for the galaxy internal dynamics, and it will not be used through  the main part of this paper.  In the Appendix \ref{appendixA} we numerically solve the variation of $\Lambda$ inside the galaxies that are analyzed in this paper (NGC 4494 and NGC 4374) and show that albeit the value of $\Lambda$ can increase many times inside a galaxy, it is far from sufficient to lead to any significative observational effect. To our knowledge, this is the first time that the variation of $\Lambda$ in galaxies was directly evaluated  in the context of renormalization group effects.

There is a simple procedure to map the solutions from the Einstein equations with the gravitational constant $G_0$  into RGGR solutions. In this review, we will proceed to find RGGR solutions via a conformal transformation of the Einstein-Hilbert action, and to this end first we write 
\be
	G = G_0 + \delta G,
\ee 
and we assume $\delta G / G_0 \ll 1$, which will be justified latter. Introducing the conformally related metric
\be
	\bar g_{\mu \nu} \equiv \frac {G_0}{G} g_{\mu \nu}, 
	\label{ct}
\ee
the RGGR action can be written as
\be
	S_{\mbox{\tiny RGGR}}[g] = S_{\mbox{\tiny EH}}[\bar g] + O(\delta G^2),
\ee
where $S_{\mbox{\tiny EH}}$ is the Einstein-Hilbert action with $G_0$ as the gravitational constant. The above suggest that the RGGR solutions can be generated from the Einstein equations solutions via the conformal transformation (\ref{ct}). Indeed, within a good approximation, one can check that this relation persists when comparing the RGGR equations of motion to the Einstein equations even in the presence of matter \cite{Rodrigues:2009vf}.

In the context of  galaxy kinematics, standard General Relativity gives essentially the same predictions of Newtonian gravity. In the weak field limit and for velocities much lower than that of light, the gravitational dynamics can be derived from the Newtonian potential, which is related to the metric by 
\be
	\bar g_{00} = - \left ( 1  + \frac {2 \Phi_\Newt}{c^2} \right ).
\ee
Hence, using eq. (\ref{ct}), the effective RGGR potential $\Phi$ is given by
\be
	\Phi = \Phi_\Newt + \frac {c^2}2 \frac{\delta G}{G_0}.
	\label{PhiRGGR}
\ee
An equivalent result can also be found from the geodesics  of a test particle\cite{Rodrigues:2009vf}. For weak gravitational fields $\Phi_\Newt/ c^2 \ll 1$ (with $\Phi_\Newt = 0$ at spatial infinity), hence even if  $\delta G/G_0 \ll 1$ eq. (\ref{PhiRGGR}) can lead to a significant departure from Newtonian gravity. 

In order to derive a test particle acceleration, we have to specify the proper energy scale $\mu$ for the problem setting in question, which is a time-independent gravitational phenomena in the weak field limit. This is a recent area of exploration of the renormalization group application, where the usual procedures for high energy scattering of particles cannot be applied straightforwardly. Previously to \cite{Rodrigues:2009vf} the selection of $\mu \propto 1/r$, where $r$ is the distance from a massive point, was repeatedly used, e.g. \cite{Reuter:2004nv,Dalvit:1994gf,Bertolami:1993mh,Goldman:1992qs, Shapiro:2004ch}. This identification adds a constant velocity proportional to $\nu$ to any rotation curve. Although it was pointed as an advantage due to the generation of ``flat rotation curves'' for galaxies, it introduced difficulties with the Tully-Fisher law \cite{Tully:1977fu}, the Newtonian limit, and the behavior of the galaxy rotation curve close to the galactic center, since there the behavior is closer to the expected one without dark matter. In \cite{Rodrigues:2009vf} we introduced a $\mu$ identification that seems better justified both from the theoretical and observational points of view. The characteristic weak-field gravitational energy scale does not comes from the geometric scaling $1/r$, but should be found from the Newtonian potential $\Phi_\Newt$, the latter is the field that characterizes gravity in such limit. Therefore,
\be
	\frac{\mu}{\mu_0} = f\( \frac{\Phi_N}{\Phi_0}\).
	\label{murggr}
\ee
If $f$ would be a complicated function with dependence on diverse constants, that would lead to a theory with small (or null)  prediction power. The simplest assumption, $ \mu \propto \Phi_\Newt$,  leads to $\mu \propto 1/r$ in the large $r$ limit; which is unsatisfactory on observational grounds (bad Newtonian limit and correspondence to the Tully-Fisher law). One way to recover the Newtonian limit is to impose a suitable cut-off, but this rough procedure does not solves the Tully-Fisher issues \cite{Shapiro:2004ch}. Another one is to use \cite{Rodrigues:2009vf}
\be
	\frac {\mu}{\mu_0} =\left( \frac{\Phi_\Newt}{\Phi_0} \right)^\alpha,
	\label{muphi}
\ee
where $\Phi_0$ and $\alpha$ are constants. Apart from the condition $ \Phi_0 < 0$, in order to guarantee $\delta G/G_0 \ll 1$, the precise value of $\Phi_0$ is largely irrelevant for the dynamics, since $\Phi'(r)$ does not depends on $\Phi_0$. The relevant parameter is $\alpha$, which will be commented below.  The above energy scale setting (\ref{muphi}) was recently re-obtained from a more fundamental perspective \cite{Domazet:2010bk}, where a renormalization group scale-setting formalism is employed.

The parameter $\alpha$ is a phenomenological parameter that needs to depend on the mass of the system, and it must go to zero when the mass of the system goes to zero. This is necessary to have a good Newtonian limit. From the Tully-Fisher law, it is expected to increase monotonically with the increase of the mass of disk galaxies. In a recent paper, an upper bound on $\nu \alpha$ in the Solar System was derived \cite{Farina:2011me}. In galaxy systems, $\nu \alpha|_{\mbox{\tiny Galaxy}} \sim 10^{-7}$, while for the Solar System, whose mass is about $10^{-10}$ of that of a galaxy,  $\nu \alpha|_{\mbox{\tiny Solar System}} \lesssim 10^{-17}$. It shows that a linear increase on $ \alpha$ with the mass (ignoring possible dependences on the mass distribution) is sufficient to satisfy both the current upper bound from the Solar System and the results from galaxies. Actually, in Sec. \ref{Sec.RoleOfAlpha} it is shown that a close-to-linear dependence on the mass can also be found for elliptical galaxies by using the fundamental plane.

Once the $\mu$ identification is set, it is straightforward to find the rotation velocity for a static gravitational system sustained by its centripetal acceleration \cite{Rodrigues:2009vf},
\be
	V^2_{\mbox{\tiny RGGR}} \approx V^2_\Newt \left ( 1 - \frac {\nu \, \alpha  \, c^2} {\Phi_\Newt} \right ).
	\label{v2rggr}
\ee
Contrary to Newtonian gravity, the value of the Newtonian 
potential at a given point does play a significant role 
in this approach. This sounds odd from the perspective of 
Newtonian gravity, but this is not so 
from the General Relativity viewpoint, since the latter has no 
free zero point of energy. In particular, the Schwarzschild 
solution is not invariant under a constant shift of the 
potential.

Equation (\ref{v2rggr}) was essential for the derivation of  galaxy rotation curves. Since elliptical galaxies are mainly supported by velocity dispersions (VD), the main equation for galaxy  kinematics in this case will not be eq.(\ref{v2rggr}), but eq.(\ref{sigma_pK}) with the mass $M(r)$ given by eq. (\ref{MRGGR}).

\section{Mass modeling of elliptical galaxies}

While disk galaxies are extended gravitational systems mainly supported by rotation, elliptical galaxies are mainly supported by velocity dispersions (VD).  There is good evidence that the elliptical galaxies dealt in this paper (and many others) are close to spherical systems, and we will consider this approximation in this paper.

For stationary spherical systems without rotation, the mean velocity at a small cell centered at position $\mathbf{r}$ is $\< \mathbf{v} \>(\mathbf{r}) = \mathbf{0}$, while the mean square velocity satisfies $\< v^2_\theta\> = \< v^2_\varphi\>  $, where $\theta$ and $\varphi$ refer to the angular components of spherical coordinates. For such systems, from  the colissionless Boltzman equation one derives the following Jeans equation \cite{0691084459},
\be
	\frac1{\ell} \frac{\prt}{\prt r} \( \ell \sigma^2_r\) + \frac 2 r \beta \sigma^2_r =  -\frac{\prt \Phi}{\prt r}.
	\label{jeans}
\ee
In the above, $\sigma^2_r (r)\equiv \<v^2_r \>(r) - \< v_r\>^2(r) = \<v^2_r \>(r) $ is the VD radial component  at the radius r, $\ell$ is the local luminosity density, $\beta(r) \equiv 1 - \sigma^2_\theta(r) / \sigma^2_r(r)$ is the local anisotropy and $\Phi$ is the potential of the total  force per mass that acts in the cell centered at $\mathbf{r}$ (i.e., $- \mathbf{\nabla} \Phi(r) = \ddot{\mathbf{r}}$). In the absence of dark matter, and considering Newtonian gravity, $\Phi$ would be the gravitational potential generated by the stars alone (we consider only elliptical galaxies with negligible amount of gas).

It is a well known theorem of Newtonian  gravity that for spherical systems 
\be
	\frac{\prt \Phi}{\prt r} = \frac{G_0 M(r)}{r^2},
\ee
where $G_0$ is the Newton's constant and $M(r)$ the total mass inside the radius $r$. Therefore the effect of a spherical dark matter profile  is to replace the total mass from the stellar one $M_*(r)$ to $M(r) = M_*(r) + M_{dm}(r)$. Hence, from the knowledge of $M_*(r)$, $M_{dm}(r)$, $\ell(r)$ and $\beta(r)$, one can solve the Jeans eq. (\ref{jeans}) to find the radial VD contribution induced by the stellar mass  ($\sigma^2_{r *}$), the radial VD contribution from the dark matter  ($\sigma^2_{r \,dm}$), and the total radial VD, which satisfies 
\be
	\sigma^2_{r } = \sigma^2_{r *}+\sigma^2_{r  \, dm}.
	\label{sigmardecomp}
\ee 
The above decomposition of $\sigma_r$ is a consequence of $\Phi$ being linear on the mass and the left hand side of eq. (\ref{jeans}) being linear on $\sigma_r^2$. The case of non-Newtonian gravity will be commented latter.

Astrophysical observations cannot measure $\sigma_r^2$ since the VD seen from the Solar System is not the radial, but the line of sight VD (which we will also call ``projected VD''). From straightforward geometrical considerations, the observed projected VD can be computed from \cite{0691084459} 

\ba
	I(R) &=& 2 \int_R^\infty \frac{\ell(r) \, r \, dr}{\sqrt{r^2 - R^2}},\\[.1in]
	\label{IandEll}
	\sigma_p^2 (R)&=&  \frac 2 {I(R)}  \int_R^\infty \( 1 - \beta(r) \frac{R^2}{r^2}\)\frac{ \ell(r) \, \sigma_r^2(r) \, r \, dr }{\sqrt{r^2 - R^2}}.
	\label{sigma_p}
\ea

The procedure described above to first evaluate $\sigma_r^2$ to then find $\sigma_p^2$ is computationally demanding (specially for models with constant but free $\beta$). A significant computational time improvement is achieved by bypassing the computation of $\sigma_r^2$ \cite{Mamon:2004xk},
\be
	 \sigma_p^2(R) = \frac {2 G_0}{I(R)} \int_R^\infty K\(\frac r R\) \frac{\ell(r) M(r) } r dr,
	 \label{sigma_pK}
\ee
where
{\small
\be
	K(u) \equiv \frac 12 u^{2 \beta - 1}\[ \(\frac 32 - \beta \) \sqrt \pi \frac{\Gamma(\beta - 1/2)}{\Gamma(\beta)} + \beta B\(\frac 1{u^2}, \beta + \frac 12, \frac 12 \) -  B\(\frac 1{u^2}, \beta - \frac 12, \frac 12 \)\],
\ee}
$B(x,a,b) = \int_0^x t^{a-1} (1-t^{b-1}) dt$ is the incomplete beta function and $\Gamma$ is the Gamma function. 

\bigskip

The above concludes a short review on how to relate mass distribution to observed velocity dispersions for Newtonian gravity. Nevertheless, almost all the above can be applied to non-Newtonian gravity as well, the single exception being the relation between $\Phi$ and $M(r)$. A useful procedure to generalize the picture above to other gravity theories, is to introduce the concept of total effective mass $M_{E}(r)$ as follows,
\be
	M_{E}(r) \equiv \frac{\Phi'(r) \, r^2}{ G_0} =  \frac{g(r) \, r^2}{G_0},
\ee
where $g(r)$ is the norm of the local acceleration.

As above discussed, the  dark matter effect  is  to enhance the total mass inside a radius r from that computed from stars alone ($M_*(r)$)  to $M(r) = M_*(r) + M_{dm}(r)$. In the case of RGGR and MOND (without  dark matter), a similar phenomenology is achieved since the  potential $\Phi$ is enhanced due to a change on the gravitational theory. The latter can be interpreted as the effect of an additional effective mass such that the total effective mass is 
\be
	M_{E}(r) = M_*(r) + M_{\mbox{\tiny RGGR}}(r),
	\label{MeRGGR}
\ee
or, in the case of MOND, $M_{E}(r) = M_*(r) + M_{\mbox{\tiny MOND}}(r)$.
\bigskip

For MOND, the acceleration $ g$ felt by a test particle due to the gravitational force of a spherical mass distribution is only equal to the Newtonian one for sufficiently large accelerations ($ g \gg a_0$). In this framework, the physical acceleration $g$ is related to the Newtonian one ($g_\Newt$) by
\be
	 g \, \mu\(\frac{g}{a_0} \) =  g_\Newt.
	\label{gMOND}
\ee
The function $\mu$ (MOND's interpolating function) is such that $\mu(x) =1 $ for $x \ll 1$ and  $\mu(x) =x $ for $x \gg 1$ \cite{1983ApJ...270..371M,1983ApJ...270..365M}. This introduces a certain ambiguity to this framework, since each choice of the function $\mu$ leads to physically different results. Fortunately, for many sounding choices of $\mu$, and for the purposes of some physical tests, the differences are negligible. For concreteness, we will fix $\mu$ as one of the most cited interpolating functions, namely  the one proposed in Ref. \cite{Famaey:2005fd} which reads
\be
	\mu(x) = \frac{x}{1 + x}.
	\label{simplemu}
\ee
With the above, the eq. (\ref{gMOND}) can be explicitly solved and it leads to the following additional effective mass contribution,
\be
	M_\MOND(r) \equiv M_{E}(r) - M_*(r) = \frac 12 M_*(r) \sqrt{1 + \frac{4 \, a_0 \,  r^2} {G_0 M_*(r)} } - \frac{M_*(r)}2.
	\label{MMOND}
\ee
The constant $a_0$ is assumed to be an universal constant for MOND (i.e., all galaxies should be subjected to the physical effects induced by the same value of  $a_0$).  From a best fit to the kinematics of diverse disk galaxies, and considering the $\mu(x)$ function above, its value was found to be $a_0 = 1.35 \times 10^{-8} cm/s^2$\cite{Famaey:2006iq}. --- If we had adopted the original $\mu$ function proposed in  \cite{1983ApJ...270..371M,1983ApJ...270..365M}, whose $a_0$ value is slightly lower according to disk galaxies data \cite{Famaey:2006iq}, a worse concordance with the NGC 4374 galaxy would be found.

\bigskip

In the RGGR case, in order to find its effective additional mass, we start from the expression for the acceleration of a test particle  in the weak field and low velocity limits, which was found in \cite{Rodrigues:2009vf},\footnote{In the presence of matter as a fluid this relation is not valid as an equality, but within a good approximation (since the $G$ variation is very small), the relation (\ref{PhiRGGR2}) holds. See Ref. \cite{Rodrigues:2009vf} for further details. }
\be
	\Phi'_\RGGR \approx \Phi'_N \( 1 - \frac {c^2  \alpha \nu}{\Phi_N} \),
	\label{PhiRGGR2}
\ee
where $\Phi_N$ stands for the Newtonian potential of all the relevant matter. In the case of elliptical galaxies, assuming no dark matter, it is essentially the Newtonian potential from the stars (i.e., $\Phi_N = \Phi_*$).

Equation (\ref{PhiRGGR2}) is the analogous of the eqs. (\ref{gMOND}, \ref{simplemu}), since it establishes the relation between the physical acceleration (considering its theoretical framework) to the Newtonian one. Contrary to MOND, eq. (\ref{PhiRGGR2}) poses no fundamental acceleration scale. In particular, there is no MOND's $\mu(x)$ interpolating function that is capable to lead to the relation given in eq.(\ref{PhiRGGR2}).

For spherical systems, 
\ba
	\Phi'_\RGGR(r) &=& \frac{G_0 M_*(r)}{r^2} \( 1 + \frac{c^2 \alpha \nu}{\frac{G_0 M_*(r)}{r} + 4 \pi G_0 \int_r^{\infty } \rho_*(r') r' dr' }\) \\
	&=&  \frac{G_0 M_*(r)}{r^2} + \frac{c^2  \alpha \nu}{r + \frac{4 \pi r^2}{M_*(r)}\int_r^{\infty } \rho_*(r') r' dr' }.
\ea
Hence, the effective additional mass of RGGR for spherical systems is given by 
\be
	M_{\mbox{\tiny RGGR}} (r) \equiv (\Phi'_\RGGR - \Phi'_N) \frac {r^2}{G_0} = \frac{\alpha \nu c^2}{G_0} \frac{r}{1 + \frac{4 \pi r }{M_*(r)} \int_{r}^{\infty} \rho_*(r') r' dr'}.
	\label{MRGGR}
\ee

\section{Observational data and numerical procedures}

\subsection{Observational data}

In order to disclose the RGGR consequences to real elliptical galaxies, we selected two elliptical galaxies, an ordinary (NGC 4494) and a giant one (NGC 4374). The analysis of both of them are based on recent data, which in particular include extended line-of-sight velocity dispersion (VD) data obtained by both long-slit observations (inner radii) and by  the VD of Planetary Nebulae (PNe).

\begin{table}[htdp]
\begin{center}
{\footnotesize
\begin{tabular}{l c c c c c c c}
\multicolumn{8}{c}{\emph{ \normalsize Distance, luminosity, expected $\Upsilon_*$ and S\'ersic parameters}}\\
\hline \hline
Galaxy $^{(1)}$		&  Distance$^{(2)}$	&$L_{\mbox{\tiny V}}$$^{(3)}$	&$\< \Upsilon_*\>^{\; (4)}$		&$R_e$$^{\; (5)}$	&$n^{(6)}$	&S\'ersic Ext.$^{(7)}$	&  Main Ref.'s$^{(8)}$ \\ 
				&  (Mpc)			&($10^{10} L_{\odot_V}$)		& ($M_\odot / L_{\odot_V}$)	& 				& 			&		 			& \\ \cline{1-8}
NGC 4374 		&	17.1 			&7.64 					&$ 4.5 $					&113.5'' 			&6.11		&290''		  	 	& \cite{2009ApJS..182..216K, 2009MNRAS.394.1249C, 2011MNRAS.411.2035N}\\
NGC 4494 		& 15.8 			& 2.64					&$ 3.8 $					& 48.2''			&3.30		& 273''				& \cite{ 2009MNRAS.394.1249C,2009MNRAS.393..329N} \\
\hline \hline
\end{tabular}
\caption {\label{DLS} \footnotesize  3)  Luminosity in the $V$ band. 4) The expected V-band stellar mass-to-light ratio considering a Kroupa IMF, see comments in text.  5)  The effective radius. 6)  The S\'ersic index. The last two values were found from S\'ersic profile fits to the observational surface brightness along the projected intermediate radius (for details see \cite{2011MNRAS.411.2035N, 2009MNRAS.393..329N}). 7) The radius at which the S\'ersic extension to the observational surface brightness is implemented here.}}
\end{center}
\end{table}%

In Table \ref{DLS} we list the main global properties of each galaxy which were relevant for the evaluation of their mass models. The expected stellar mass-to-light ratio is based on previous analysis that use the Kroupa IMF \cite{2001MNRAS.322..231K, 2002Sci...295...82K}, see also Ref. \cite{2012ApJ...748....2D}. We have not added to the table above, but  $\beta$ estimates can  be found in Ref. \cite{2012ApJ...748....2D}, where it is found that both of the galaxies are close to isotropic ($-0.5 <\beta < 0.5$). Converting to the V band the $\Upsilon_*$'s derived in Refs. \cite{Gerhard:2000ck, 2009MNRAS.396.1132T}, it is fair to assume that, for NGC 4374, $\<\Upsilon_*\>_{\mbox{\tiny Kroupa IMF}} \sim (4.5 \pm 1.0) M_\odot/L_{\odot, V}$ (see also Ref. \cite{2011MNRAS.411.2035N}). For NGC 4494, we use $\<\Upsilon_*\>_{\mbox{\tiny Kroupa IMF}} \sim (3.8 \pm 1.0) M_\odot/L_{\odot, V}$ \cite{2009MNRAS.393..329N}  (it should be noted that this galaxy kinematics is not compatible with the Salpeter IMF  \cite{2012ApJ...748....2D}). It is not our purpose to be ultimately precise on the Kroupa IMF, the essential issue is to verify that RGGR has a tendency of being in conformity with IMF's that lead to significantly lower $\Upsilon_*$'s than the Salpeter one. This tendency could thus provide another  important physical test for our proposal once stelar population models become more precise.

For each of the above galaxies, we use the observational surface brightness data up to a the largest radius where the observation is trustworthy, and then  extend it with its S\'ersic profile  (analogously to the case of disk galaxies, where the extension is implemented with a simple exponential profile).  The observational surface brightness profile, which we call $\mu_{\mbox{\tiny obs}}(R)$, is determined from observations up to a certain radius. It is currently known that the surface brightness of many galaxies can be  well approximated by  S\'ersic profiles \cite{1968adga.book.....S}. From $\mu_{\mbox{\tiny obs}}$ one can fit the parameters $R_e$ and $n$ to find the corresponding S\'ersic surface brightness of each galaxy, which we call $\mu_{\mbox{\tiny S}}(R)$. The latter can be used to generate an extended surface brightness profile $\mu_{\mbox{\tiny ext}}$, which is composed by $\mu_{\mbox{\tiny obs}}$ from the center up to a given radius and $\mu_{\mbox{\tiny S}}$ from that point out to a larger radius. 

The intensity associated with the S\'ersic profile  reads (see Ref. \cite{2005PASA...22..118G} for a review),
\be
	I(R) = I_0 \, e^{- \beta_n \( \frac R {R_e}\)^{1/n}},
	\label{sersicprof}
\ee
where $R$ is the projected radius, $n$ is the S\'ersic index and $R_e$ is the effective radius (which is defined as the radius that encloses half of the total luminosity). Since the total luminosity is given by $L = 2 \pi \int_0^\infty I(R) dR^2$,  the constant $\beta_n$  must satisfy $\Gamma(2 n) = 2 \gamma(2 n , \beta_n)$, where $\Gamma$ and $\gamma$ stand for the gamma function and the (generalized) incomplete gamma function (i.e., $\gamma(a,b) = \int_0^b t^{a - 1} e^{-t} dt$). Solutions for $\beta_n$ can be found numerically.

Since most photometric observations are stated in unities of mag/arcsec$^2$,  the following relation between surface brightness $\mu(R)$ and the intensity $I(R)$ is useful,
\be
	\mu(R) = - 2.5 \, \mbox{log}_{10} \[ I(R) \frac{  \mbox{kpc}^2}{L_\odot  } \( \frac { \pi 10^{-2}}{180 \times 60 \times 60}\)^2   \]  + {\cal M}_\odot,
\ee
where ${\cal M}_\odot$ is the magnitude of the sun in the appropriate band.

\bigskip

\subsection{The stellar contribution to the observed VD} \label{stelarVD.subsec}

To find the stellar contribution to the total projected VD, we proceed as follows: \\

\noindent
$i$)  Find the luminosity density, which comes from the inversion of  eq. (\ref{IandEll}) \cite{0691084459}, namely
\be
	\ell (r)= - \frac{1} {\pi} \int_r^\infty \frac{I'(R)}{\sqrt{R^2 - r^2}} dR.
\ee
From the  above,  a vector $\{\ell(r_i)\}$  is built. This vector is numerically interpolated to generate a numerical function for $\ell(r)$ (this step is necessary to improve computational performance). At this and similar procedures, the $r_i$ resolution is chosen such that  the errors between the  original function and the one build from the interpolation are no more  than $1 \%$. \\ 

\noindent
$ii$) Using a linear relation between stellar mass density and luminosity density, $\rho_* (r)= \Upsilon_* \, \ell(r)$, it is straightforward to build the total stellar mass inside the radius $r$, $M_*(r)$. \\

\noindent
$iii$) We use $\ell$, $M_*$ and the eq. (\ref{sigma_pK}) to build the numerical table $\{\sigma^2_{*} (R_i, \beta_j)/ \Upsilon_*\}$, with $\beta_j \in [-1,1]$. In general, $\beta \in (- \infty, 1]$, but   physical considerations (based  on galaxy formation or reasonable restrictions on the distribution function) disfavor large negative (tangential) anisotropy, while favors either isotropy or radial anisotropy \cite{2011MNRAS.411.2035N,2009MNRAS.393..329N}.

\bigskip

The three steps above determines $\sigma_{p*}$, apart from  $\Upsilon_*$ and $\beta$, which will be  analyzed latter.

\bigskip

\subsection{The non-Newtonian  contributions to the observed VD}

To find the RGGR additional contribution to the projected VD, which we call $\sigma_{p \, \RGGR}(R)$ and satisfies 
\be
	\sigma_p^2 = \sigma^2_{p *} + \sigma^2_{p\, \RGGR},
	\label{sigmap}
\ee
 we  re-do the steps $ii$ and $iii$ from above with  $M_\RGGR$ (see eq. (\ref{MRGGR})) in place of $M_*$. With this procedure, the total projected VD for RGGR is completely determined except for $\beta$ and two constants that appear linearly inside $\sigma_{p *}^2$ and $\sigma_{p \, \RGGR}^2$, namely: the stellar mass-to-light ratio $\Upsilon_*$ and the constant $\nu \alpha$.

\bigskip

The effective additional mass provided by MOND (eq. \ref{MMOND}) has a  simple form, but it depends on different powers of  $\Upsilon_*$.  Hence, in the step $iii$ described previously, the numerical function $\sigma_{ p \, \MOND}$ is found from the three dimensional interpolation of the following  table, $\{ \sigma^2_{ p\, \MOND} (\Upsilon_{* i}, R_j, \beta_k ) \}$, where $\Upsilon_*$ was constrained to be in the range\footnote{No physical issues are expected due to this constrain for this galaxy in this band.} $1\le \Upsilon_*/(M_\odot/L_{\odot_V}) \le 10$, while $R$ and $\beta$ have the same resolutions and range as in the RGGR model described above.

Before proceeding, we remark that the stellar component modeling used for MOND is identical to the modeling used for RGGR and Newtonian gravity. In Ref.\cite{2011A&A...531A.100R} it is argued in favor of the use of Jaffe profiles in the outer parts of  the baryonic mass profiles of ellipticals.  Here this route is not pursued.

\section{Elliptical galaxies within RGGR without dark matter: general aspects}

\subsection{S\'ersic profiles} \label{generalaspects}

\begin{figure}[thbp]
\begin{center}
	  \includegraphics[width=100mm]{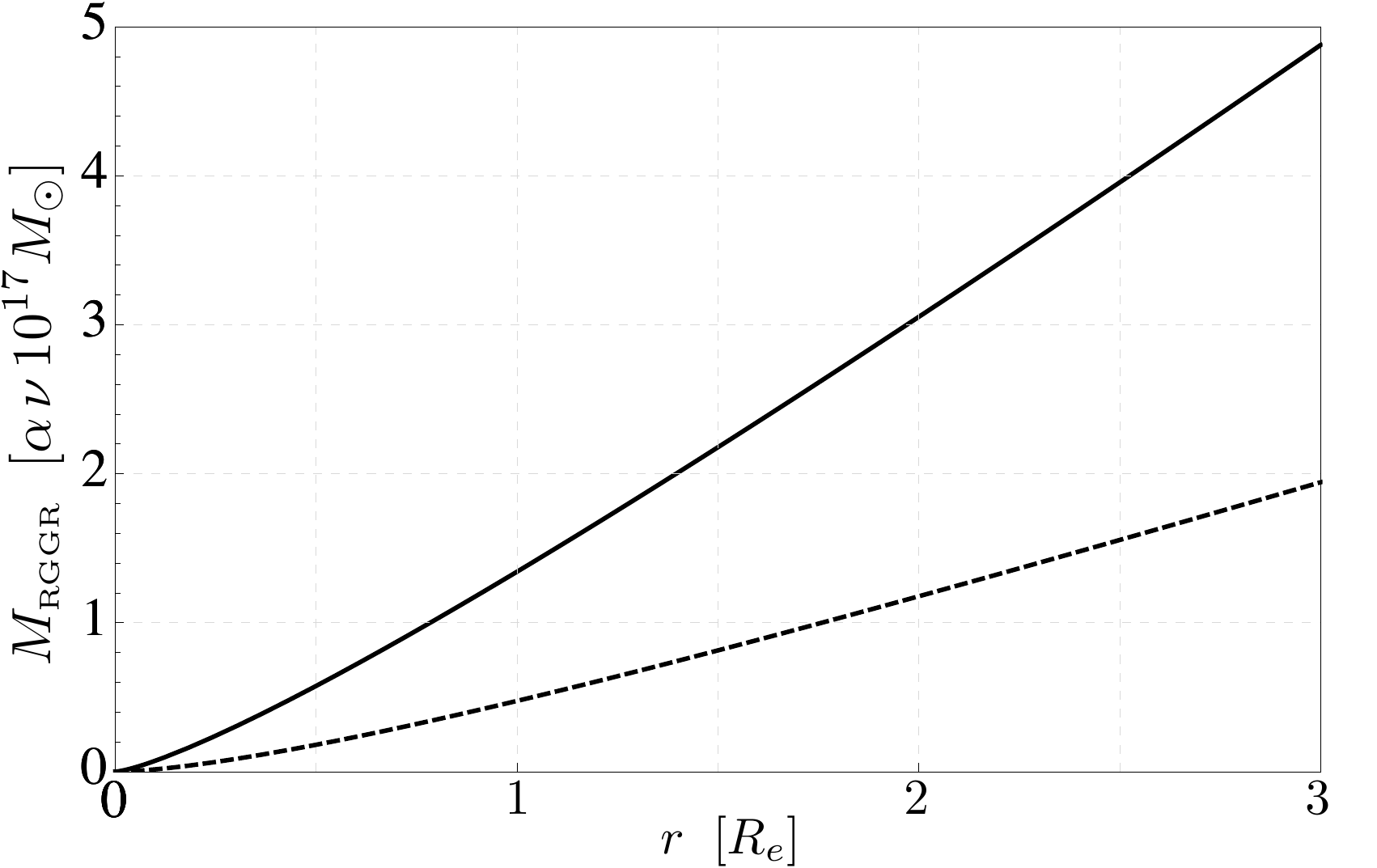}
\caption{The RGGR additional mass contribution $M_\RGGR$, as a function of the deprojected radius $r$, for two S\'ersic profiles. One of the profiles has S\'ersic index $n = 6.1$ (solid line) and the other with $n = 3.3$ (dashed line). Many of the elliptical galaxies have intermediate S\'ersic indices. For $\alpha \nu \sim 10^{-7}$, the additional mass $M_\RGGR$ is about the same order of the baryonic mass in a galaxy.}
\label{MRGGRsersic}
\end{center}
\end{figure}

\begin{figure}[thbp]
\begin{center}
	  \includegraphics[width=100mm]{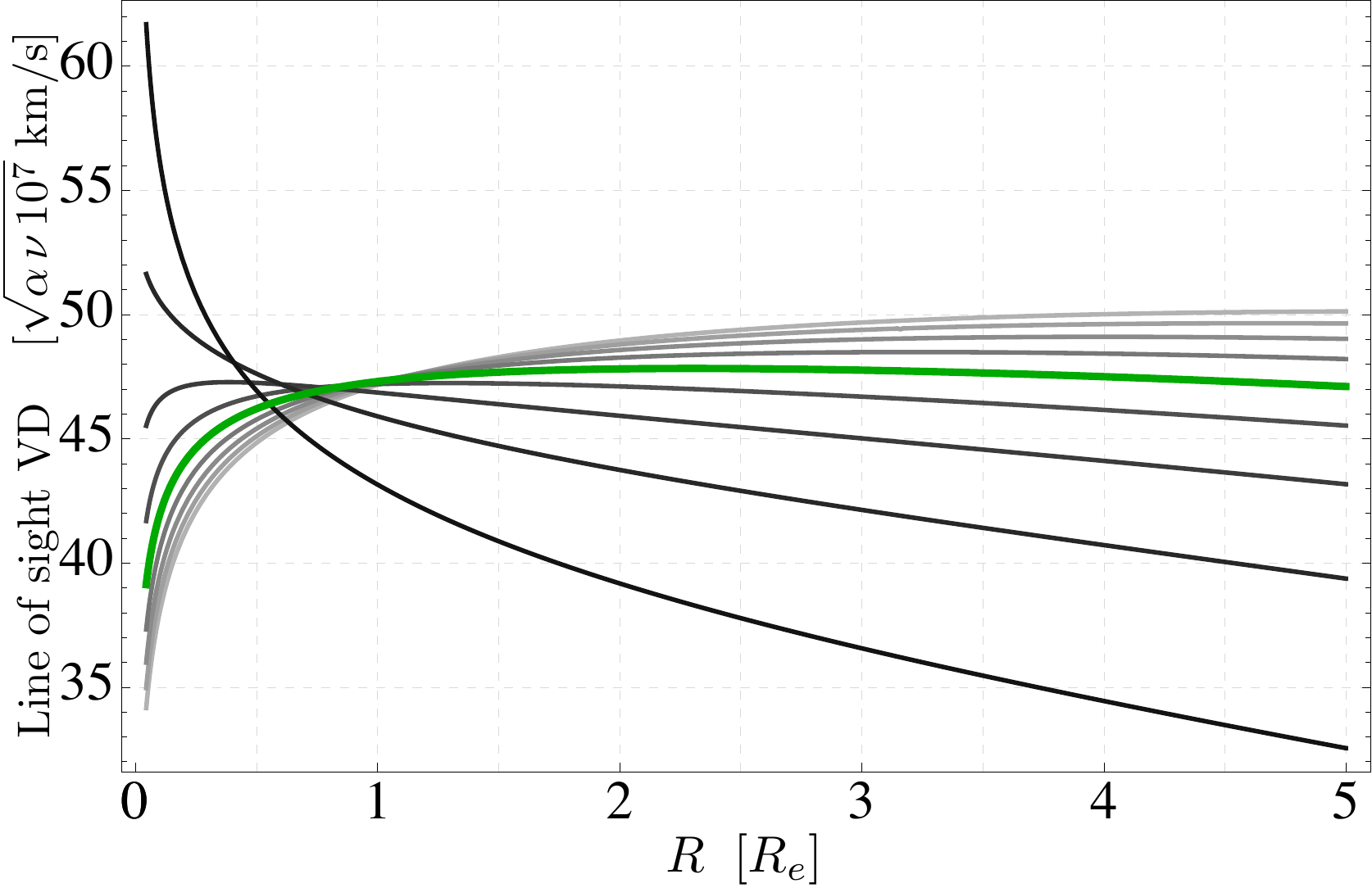}
\caption{The projected VD contribution from RGGR ($\sigma_{p \, \RGGR}(R)$) for a matter distribution given by the S\'ersic profile  (\ref{sersicprof}) with $n = 6.1$. The thickest (green) curve corresponds the  isotropic case $\beta =0 $. The others are associated to  the following $\beta$  values: -1.00, -0.75, - 0.50, - 0.25, 0.25, 0.50,  0.75, 1.00 (from light gray to black, respectively).  }
\label{SigmapRGGRSersicN4374}
\end{center}
\end{figure}

\begin{figure}[thbp]
\begin{center}
	  \includegraphics[width=100mm]{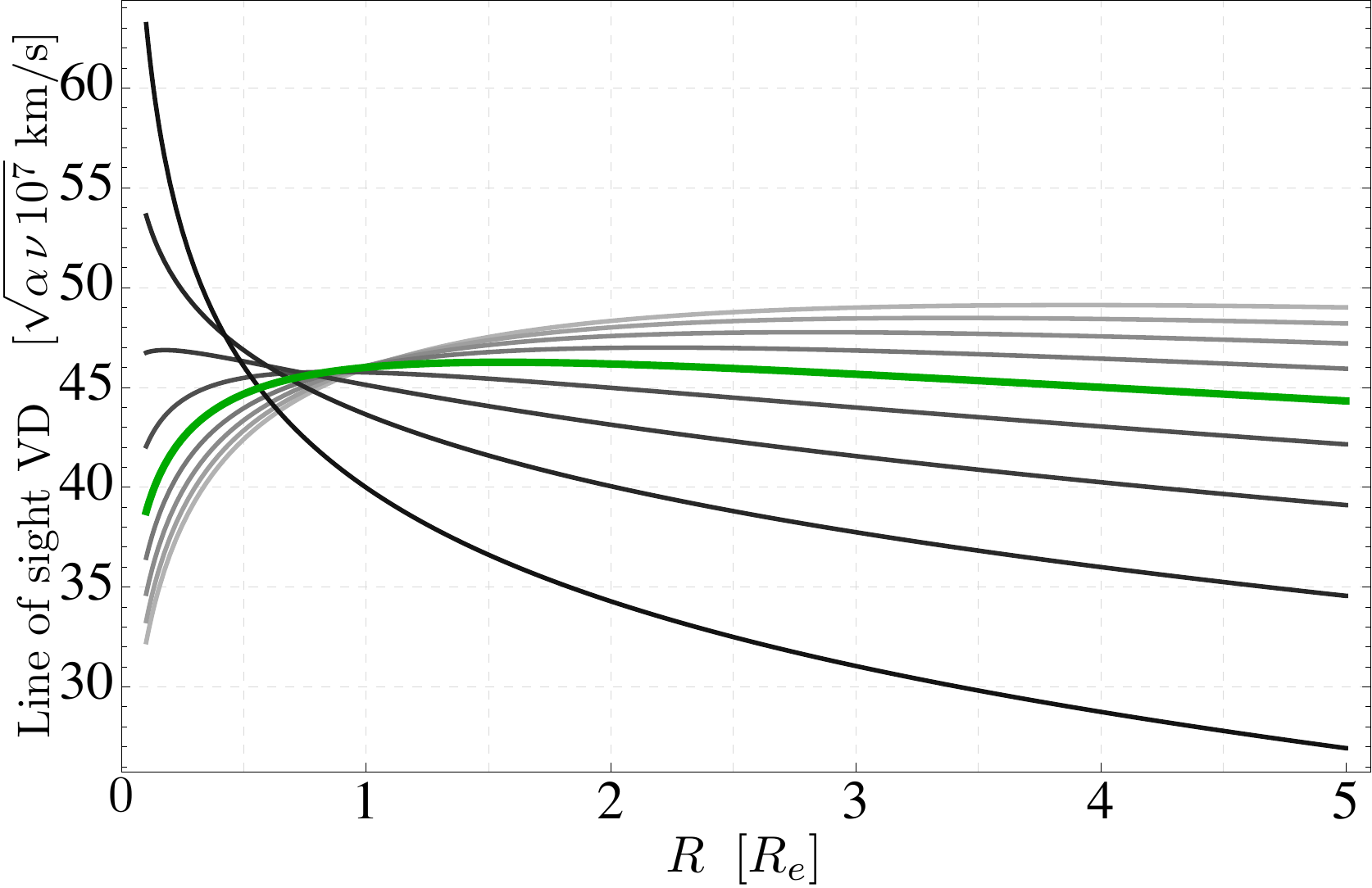}
\caption{The same of Fig. \ref{SigmapRGGRSersicN4374}, but with S\'ersic index $n= 3.3$.}
\label{SigmapRGGRSersicN4494}
\end{center}
\end{figure}

For spherical stationary systems, the main differences between Newtonian gravity and the RGGR one are encoded in the expression for $M_\RGGR$ (\ref{MRGGR}). From a given matter density $\rho(r)$,  it is straightforward to derive the additional effective mass posed by RGGR. In the end, the single uncertainty relies on the constant $\alpha$, on which further comments on its behavior will be presented in the next section.

Since the matter density of diverse elliptical galaxies are fairly proportional to a deprojected S\'ersic profile (which,  for $n=4$, coincides with the de Vaucouleurs profile), here we will consider the consequences of RGGR for matter profiles deduced from deprojected S\'ersic profiles (for all radii). Two cases will be explicitly evaluated: the case $n=3.3$ and $n = 6.1$ (the same S\'ersic indices of NGC 4374 and NGC 4494). Such values are useful to the purposes of this section, since many ellipticals have S\'ersic indices that lie between  these values.

The RGGR mass profile $M_\RGGR(r)$ is shown in Fig. \ref{MRGGRsersic}. The contributions of such profiles to the projected VD  (see eq. (\ref{sigmap})), including  different assumptions for the anisotropic parameter $\beta$, are depicted in Figs. \ref{SigmapRGGRSersicN4374}, \ref{SigmapRGGRSersicN4494}.

Apart from the  $\alpha$ role, from the eq. (\ref{MRGGR}) one sees that $M_{\RGGR}$ is not sensible to either the constant $I_0$ in eq. (\ref{sersicprof}) or the value of any constant mass-to-light ratio; it only depends on the effective radius $R_e$ and the S\'ersic index $n$.

\subsection{The role of $\alpha$ and the fundamental plane} \label{Sec.RoleOfAlpha}

Here we point and illustrate that future analysis on the relation between RGGR and the fundamental plane  \cite{1987ApJ...313...59D,1987ApJ...313...42D}   can provide significant constraints on the correlation between $\alpha$ and the mass distribution. It is beyond the scope of this work to do a solid analysis on the RGGR consequences for the fundamental plane interpretation (interpretation which is currently an open problem within any framework). In particular, the status of such task for MOND can  be checked in Ref.  \cite{2011MNRAS.412.2617C}  and references therein. 

The existence of a plane in the space $(\< I\>_e, R_e, \sigma_0)$ (the fundamental plane) is usually interpreted as a consequence of the virial theorem (in the context of Newtonian gravity), see Ref. \cite{0521857937}  for a   review. Also, the data scattering can be further reduced around the theoretical surface if the S\'ersic index $n$ is included as a new axis, a ``fundamental hyperplane'' \cite{2002MNRAS.334..859G}. It is admissible to use the same relation in context of RGGR, except that  one should replace the virial mass $M_v$ by the corresponding effective mass ${M_E}_v$ (\ref{MeRGGR}) (considering stationary and spherically symmetric systems). Note that the total effective mass is not finite, but such feature is not a novelty, since well known dark matter profiles do display the same feature. One may define, for instance, the virial radius at the radius at which the local effective density is about 200 times the cosmological critical density. The details of such definition will not be relevant to the approach below.

The theoretical plane in the $(\< I\>_e, R_e, \sigma_0)$ space, which is derived from Newtonian gravity, together with the virial theorem and with  constant $M/L$,  is tilted in regard to the observed one. The tilt can be interpreted as an evidence of dark matter (within the Newtonian gravity context), since the necessary $M/L$ variation to match theory and observation is larger than the expected from stellar population variations (e.g., \cite{2006MNRAS.366.1126C}). Hence, one way to explain the tilt of the fundamental plane is from the assumption that
\be
	\frac M L \propto L^A \< I\>_e^B,
	\label{MLAB}
\ee
where the values of $A$ and $B$ depends on the values of the parameters that determine the fundamental plane (from observations). For instance, using the values of Ref.  \cite{1996MNRAS.280..167J} one finds $A = 0.31$ and $B=0.02$ (see also Ref.\cite{0521857937}). Part of the above scaling relation would be due to systematic stellar mass-to-light ratios ($M_*/L$) variations.  For simplicity, we will here assume no systematic variation of $ M_*/L $, hence

\be
	\frac M L = \frac {M_*} L \frac M {M_*} \propto 1 + \frac{M_\RGGR}{M_*}.
	\label{MLMRGGR}
\ee

At the virial radius, it is fair to use $M_* \ll M_\RGGR$. Combining the last comment with eqs. (\ref{MLAB},\ref{MLMRGGR}),
\be
	\frac{M_\RGGR}{M_*} \propto I_0^{A+ B} R_e^{2A} g^{A+ B}(n),
	\label{MRGGRMstarsI0}
\ee
since $\<I\>_e = I_0  \, g(n)/ (2 \pi)$ and $L = I_0 \,R_e^2 \,g(n)$, where 
\be
	g(n) \equiv 2 \pi  \frac{n e^{2 \beta_n}}{{\beta_n}^{2n}} \Gamma(2 n) .
\ee

In Sec. \ref{generalaspects} it is shown that $M_\RGGR(r)/\alpha$  does not depend on $I_0$, hence from eq. (\ref{MRGGRMstarsI0}) a rough estimate on the variation of $\alpha$ with $M_*$ can be found as
\be
	\alpha \propto M_*^{1 + A + B} f(R_e,n).
\ee 
That is, for elliptical galaxies of about the same shape (i.e., with similar $R_e$ and $n$), $\alpha$ should increase with the stellar mass of the the galaxy faster than linear, but slower than quadratic. This is a quick procedure to evaluate the dependence of $\alpha$ with the galaxy parameters. The best procedure would be to use a large sample of elliptical galaxies to deduce that, and also unveil the function $f(R_e, n)$.

\section{NGC 4374 and NGC 4494 results}

\subsection{Introduction}

Here a detailed numerical analysis of NGC 4374 and NGC 4494 is done. The main results are shown in Tables \ref{N4374resultsTable}, \ref{N4374RGGRmassTable}, \ref{N4494resultsTable}, \ref{N4494RGGRmassTable}  and Figs.  \ref{N4374MONDFig},\ref{N437RGGRPhotoFig},  \ref{N4494RGGRPhoto3}. The main purpose of this section is to show that recent data on elliptical galaxies is in conformity with the RGGR model, even assuming a negligible amount of dark matter. Moreover, we  compare  our results to MOND (which also constitute an original work presented in this paper). 

Many elliptical galaxies display a certain degree of rotation. Such rotation do not appear directly in the velocity dispersion (VD) data, but it also traces the galaxy effective mass. In order to mass model such galaxies without simply neglecting the rotational (sub-dominant) component, one can insert a compensation in the VD data. The data that is here called  ``line of sight VD''   has already a compensation for rotation, whose computation was done in Refs.\cite{2009MNRAS.393..329N, 2011MNRAS.411.2035N}. We use the same data that these references use for mass modeling these galaxies.

\begin{table}[htdp] 
\begin{center}
{\footnotesize 
\begin{tabular}{l c c c c c c}
\multicolumn{6}{c}{\emph{ \normalsize NGC 4374}}\\
\hline \hline
\multicolumn{6}{c}{\emph{Newtonian gravity without dark matter} } 				  \\ 
Stellar model $^{(1)}$		& 			---			& $\beta^{\; (2)}$	 		&\ML $^{\; (3)}$			&${\chi^{2}}^{\; (4)}$	&${\chir}^{\; (5)}$\\ \cline{1-6}
$\beta_{[0]}$				&						& 0 						& 7.59$\pm$0.15 		& 140 			& 6.4   \\
$\beta_{[-1,1]}$				&						&$-1.00^{+0.11}_{-0.00}$		&  7.82$\pm$0.23		& 104			& 4.9 \\
K.IMF+$\beta_{[0]}$			&						& 0 						& 6.66$\pm$0.11 		& 292 			& 13   \\
K.IMF+$\beta_{[-1,1]}$		&						&$-1.00^{+0.37}_{-0.00}$		&  6.78$\pm$0.17		& 276			& 13 \\

								&						&						&					&				&\\
\multicolumn{6}{c}{\emph{RGGR without dark matter} } 	\\ 
Stellar model				& $\alpha\,\nu \times 10^{7}$$^{\;(6)}$& $\beta$		 			&\ML					&${\chi^{2}}$		&${\chir}$\\ \cline{1-6}
$\beta_{[0]}$				&$14.9\pm2.4$				&0 						&$4.14\pm0.57$		&21.1 			&1.0\\
$\beta_{[-1,1]}$				&$15.3^{+5.5}_{-4.7}$		&$0.12^{+0.66}_{-1.12}$ 		&$3.9^{+1.7}_{-2.4}$		&21.0			&1.1\\
K.IMF+$\beta_{[0]}$			&$13.8\pm1.4$				&0						&$4.41\pm0.28$		&21.8			&0.99\\
K.IMF+$\beta_{[-1,1]}$		&$14.0\pm1.9$				&$-0.18^{+0.43}_{-0.78}$		&$4.48\pm0.38$		&21.3			&1.0\\
						&					&							&					&				&\\
\multicolumn{6}{c}{\emph{MOND without dark matter} } 	\\ 
Stellar model				& ---		 				& $\beta$				&\ML					&${\chi^{2}}$		&${\chir}$\\ \cline{1-6}
$\beta_{[0]}$				&						&0					&$6.8\pm0.1$			&34.2			&1.6 \\
$\beta_{[-1,1]}$				&						&$-1.0^{+0.4}_{-0.0}$	&$7.2\pm0.2$			&24.9			&1.2\\
K.IMF+$\beta_{[0]}$			&						&0					&$6.1\pm0.1$			&117				&5.1\\
K.IMF+$\beta_{[-1,1]}$		&						&$0.4^{+0.2}_{-0.3}$		&$6.0\pm0.2$			&113				&5.2\\
\hline \hline

\end{tabular}
\caption{\label{N4374resultsTable} \footnotesize NGC 4374 results. (1) $\beta_{[0]}$ indicates isotropic VD, $\beta_{[-1,1]}$ indicates constant anisotropy with $\beta \in [-1,1]$, K.IMF is a reference to Kroupa IMF, and it means that the expected value of $\Upsilon_*$ was used, see Table \ref{DLS} and eq. (\ref{chiUps}). (4) and (5) are the chi squared and reduced chi squared parameters, for details see Sec.\ref{Sec.chi2}. }}
\end{center}
\end{table}

\begin{table}[htdp]
\begin{center}
{\footnotesize
\begin{tabular}{l c }
\hline \hline
Stellar Model						&$M_*/(10^{10} M_\odot)$  \\
\footnotesize (1)					& (2)			\\
\hline
$\beta_{[0]}$				&$31.7\pm8.5$\\
$\beta_{[-1,1]}$			&$30^{+17}_{-22}$\\
K.IMF+$\beta_{[0]}$		&$33.7\pm6.5$\\
K.IMF+$\beta_{[-1,1]}$	&$34.2\pm7.4$\\[0.2cm]

\hline \hline
\end{tabular}
\caption{\label{N4374RGGRmassTable} \footnotesize RGGR mass results for NGC 4374.}}
\end{center}

\end{table}%

\begin{table}[htdp]
\begin{center}
{\footnotesize
\begin{tabular}{l c c c c c c}
\multicolumn{6}{c}{\emph{ \normalsize NGC 4494}}\\
\hline \hline
\multicolumn{6}{c}{\emph{Newtonian gravity without dark matter} } 				  \\ 
Stellar model $^{(1)}$				& 			---			& $\beta^{\; (2)}$	 		&\ML $^{\; (3)}$			&${\chi^{2}}^{\; (4)}$	&${\chir}^{\; (5)}$\\ \cline{1-6}
$\beta_{[0]}$				&						& 0 						& 4.206$\pm$0.044 		& 20.6 			& 0.82  \\
$\beta_{[-1,1]}$			&						&$-0.55^{+0.26}_{-0.32}$		&$4.164\pm0.067$		&6.25			&0.26\\
K.IMF+$\beta_{[0]}$		&						& 0 						& 4.190$\pm$0.043 		& 23.6 			& 0.91  \\
K.IMF+$\beta_{[-1,1]}$		&						&$-0.56^{+0.25}_{-0.32}$		&$4.150\pm0.065$		&8.54			&0.34\\[.1in]

								&						&						&					&				&\\
\multicolumn{6}{c}{\emph{RGGR without dark matter} } 	\\ 
Stellar model						& $\alpha\,\nu/10^{-7}$$^{\;(6)}$& $\beta$		 			&\ML					&${\chi^{2}}$		&${\chir}$\\ \cline{1-6}
$\beta_{[0]}$				&$1.55^{+0.60}_{-0.58}$		&0						&$3.49\pm0.28$		&3.19			&0.13\\
$\beta_{[-1,1]}$ 			&$1.2^{+1.6}_{-1.2}$			&$-0.15^{+0.43}_{-0.55}$ 		&$3.67^{+0.51}_{-0.73}$	&2.81			&0.12\\
K.IMF+$\beta_{[0]}$		&$1.21\pm0.44$			&0						&$3.66\pm0.20$		&5.07			&0.20\\
K.IMF+$\beta_{[-1,1]}$	 	&$0.86^{+0.72}_{-0.69}$		&$-0.24^{+0.31}_{-0.42}$		&$3.80^{+0.29}_{-0.31}$	&3.08			&0.13\\[.1in]

								&						&						&					&				&\\
\hline \hline
\end{tabular}
\caption{\label{N4494resultsTable} \footnotesize NGC 4494 results. See Table \ref{N4374resultsTable} for details. }}
\end{center}
\end{table}%

\begin{table}[htdp]
\begin{center}
{\footnotesize
\begin{tabular}{l c }
\hline \hline
Stellar Model						&$M_*/(10^{10} M_\odot)$  \\
\footnotesize (1)					& (2)			\\
\hline
$\beta_{[0]}$						&$9.2 \pm 1.5$\\
$\beta_{[-1,1]}$						&$9.7^{+2.1}_{-2.7}$	\\
K.IMF + $\beta_{[0]}$				&$9.7\pm1.3$	\\
K.IMF + $\beta_{[-1,1]}$				&$10.0\pm1.6$	\\
\hline \hline
\end{tabular}
\caption{\label{N4494RGGRmassTable} \footnotesize  RGGR  mass results for NGC 4494.}}
\end{center}
\end{table}%

\begin{figure}[thbp]
\begin{center}
	  \includegraphics[width=110mm]{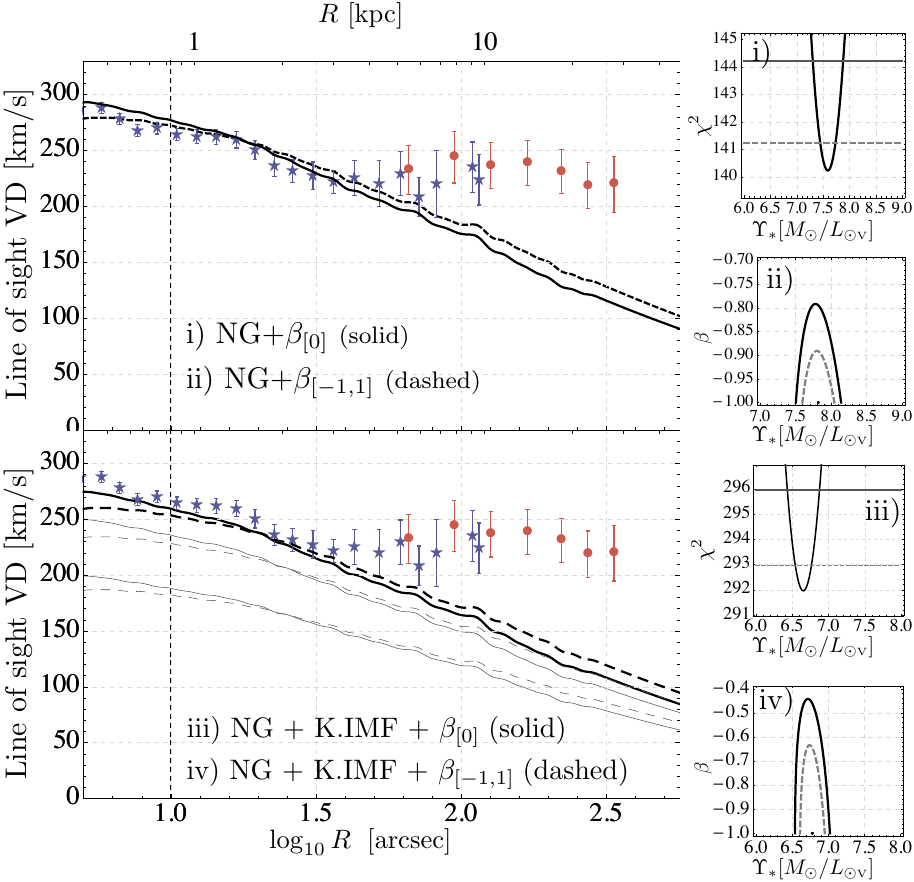}
\caption{\footnotesize NGC 4374 mass models with Newtonian gravity (NG). In the larger plots at the left, the circles and stars with error bars refer to the observational VD, either from long-slit observations (the stars) or from planetary nebulae (the circles). The curves refer to mass models composed  by a stellar component   with Newtonian gravity (NG). The first model (i) assumes isotropy ($\beta = 0$), while the second (ii) assumes $\beta \in [-1,1]$. Models iii and iv use the expected mass-to-light ratio from the Kroupa IMF (K.IMF) as part of the data to be fitted (see eq. (\ref{chiUps}) and Table \ref{DLS} for details).  The four thin lines that appear in the lower left plot depict the  theoretical error bars on  the stellar velocity dispersion (the solid thin line refers to the model iii, while the dashed thin line to the model iv, which differ since they have different $\beta$ values). The vertical dashed line signs the radius above which the observational data is considered  for the fitting procedure (10 arcsec) \cite{2011MNRAS.411.2035N}. The four small plots at the right show the $1 \sigma$ and $2 \sigma$ confidence levels for each of the models. } \label{N4374StarsPhotoFig}
\end{center}
\end{figure}

\begin{figure}[thbp]
\begin{center}
	  \includegraphics[width=140mm]{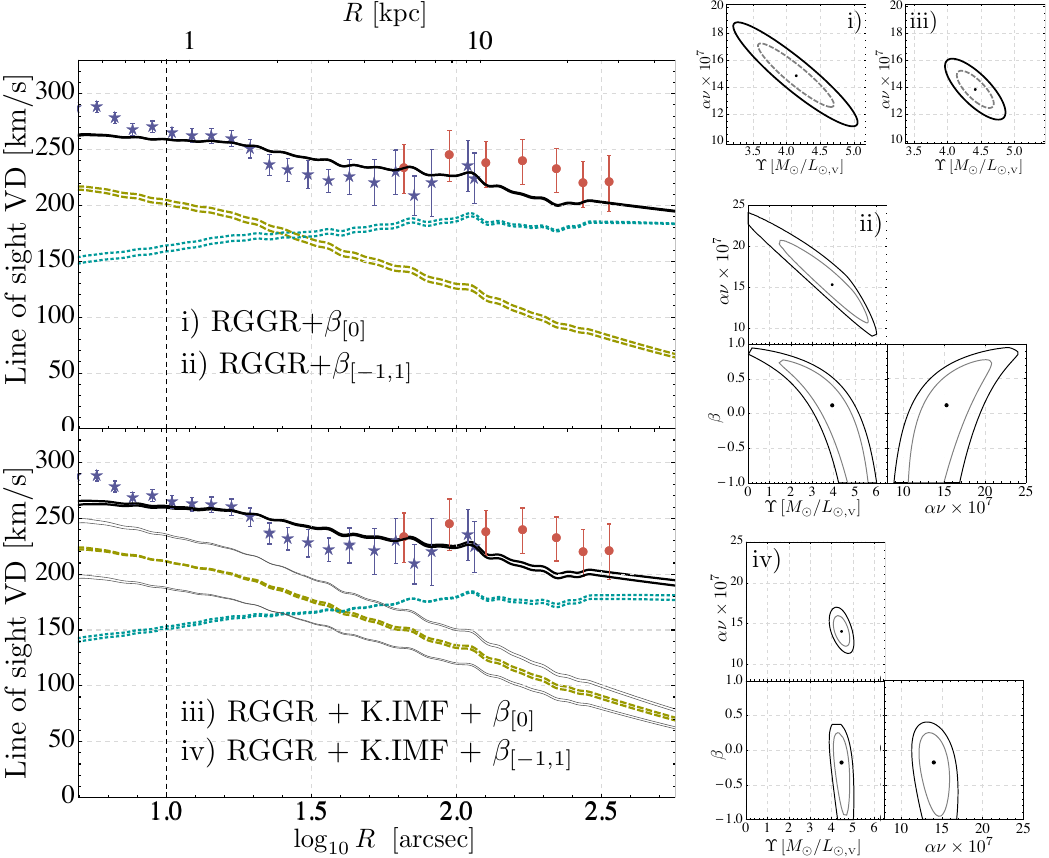}
\caption{NGC 4374 mass models with RGGR. The curves refer to mass models  composed by the stellar component (inferred from photometric data with S\'ersic extrapolation) and RGGR gravity. The black solid line in each of the four large plots is the resulting VD for each model, the yellow dashed and blue dotted lines are respectively the stellar Newtonian and non-Newtonian contributions to the total VD, and the thin dark gray lines in models iii and iv are the effective error bars for $\Upsilon_*$ (see text for details). The first model (i) assumes isotropy ($\beta = 0$),  the second (ii)  assumes $\beta \in [-1,1]$, while models iii and iv also consider the expected mass-to-light ratio ($\Upsilon_*$) in the fitting procedure  (see eq. (\ref{chiUps}) and Table \ref{DLS} for details). The vertical dashed line signs the radius above which the observational data is considered  for the fitting procedure (10 arcsec). The smaller plots at the right show the $1 \sigma$ and $2 \sigma$ confidence levels for each of the four models.}
\label{N437RGGRPhotoFig}
\end{center}
\end{figure}

\begin{figure}[thbp]
\begin{center}
	  \includegraphics[width=125mm]{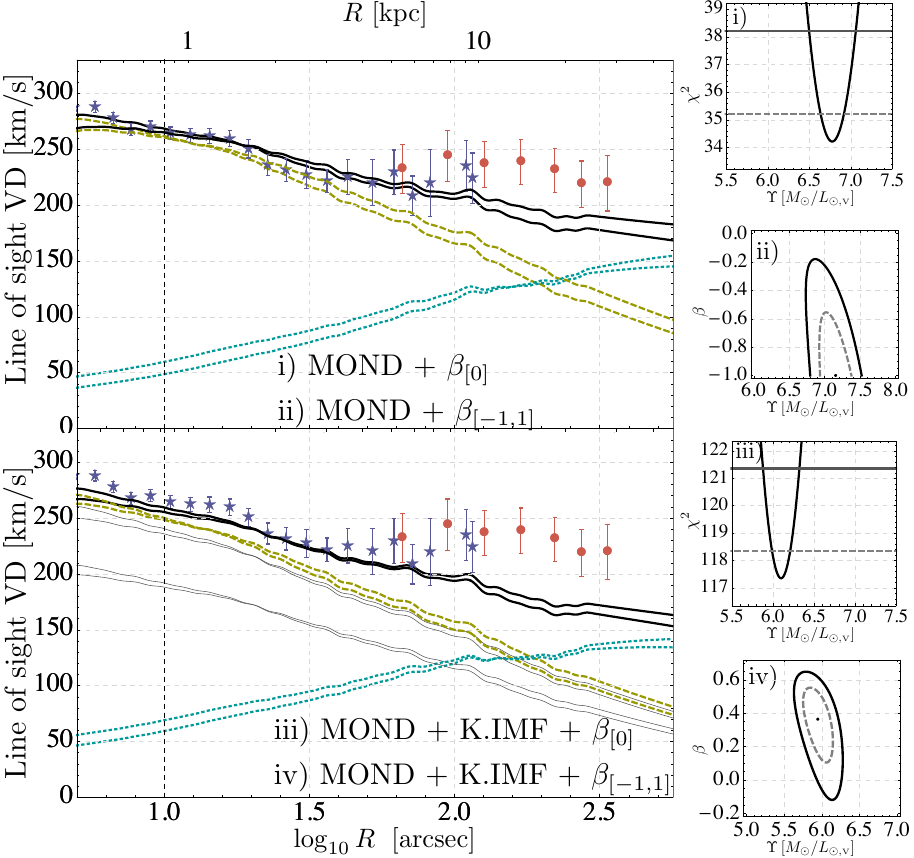}
\caption{ NGC 4374 mass models with MOND. See Fig. \ref{N437RGGRPhotoFig} for details.}
\label{N4374MONDFig}
\end{center}
\end{figure}

\begin{figure}[thbp]
\begin{center}
	  \includegraphics[width= 110mm]{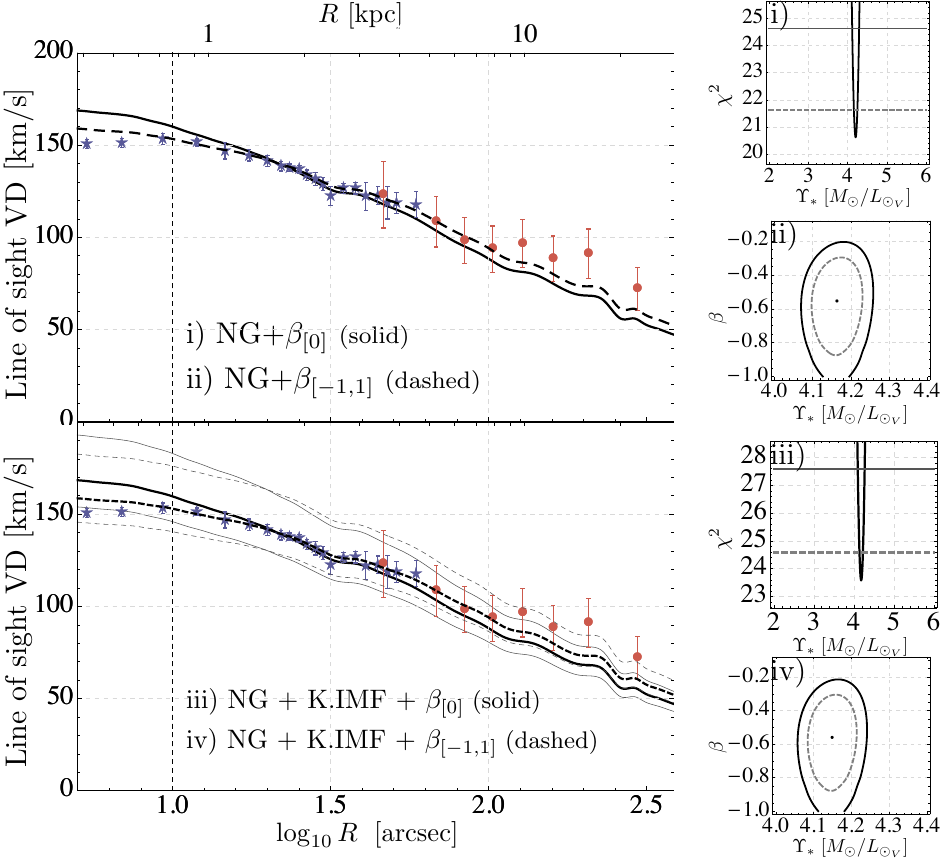}
\caption{NGC 4494 mass models with Newtonian gravity (NG). See Fig. \ref{N4374StarsPhotoFig} for details. }
\label{N4494StarsPhotoFig}
\end{center}
\end{figure}

\begin{figure}[thbp]
\begin{center}
	  \includegraphics[width=140mm]{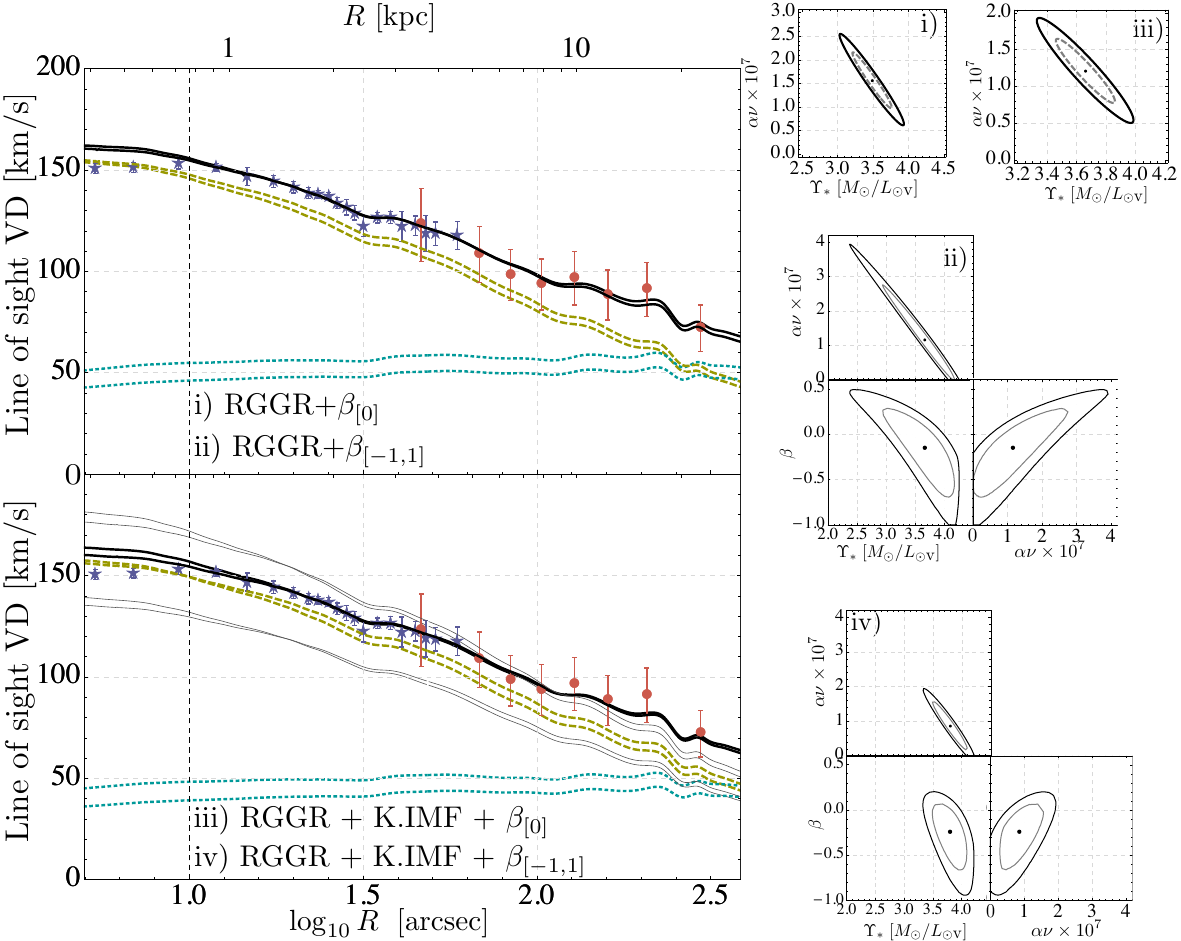}
\caption{ NGC 4494 mass models with RGGR. See Fig. \ref{N437RGGRPhotoFig} for details.}
\label{N4494RGGRPhoto3}
\end{center}
\end{figure}

In the Newtonian gravity without dark matter framework, there are two parameters that the procedures  described in Sec.\ref{stelarVD.subsec}  do not fix, namely: the mass-to-light raio $\Upsilon_*$ and the anisotropy parameter $\beta$. There are theoretical expectations for $\Upsilon_*$, but they are considerably more uncertain than other astrophysical data. 

Contrary to MOND, the RGGR approach was not developed in order to remove the necessity of dark matter, but to explore a possible QFT effect on large scales that may have a significant impact on astrophysics and cosmology. In this paper, likewise in Ref.\cite{Rodrigues:2009vf}, we explore the extremum possibility of having no dark matter in galaxies and only standard gravitation with the running of $G$ as given by eq. (\ref{gmu}). It would be a leap of faith the assume that this single running of $G$ could solve all the dark matter issues at all scales, and we do not expect that. On the other hand, the galactic picture we are describing can be more than a mere starting point, since it does work as reasonable approximation of a universe with a lower amount of  dark matter, and of a warmer type. Considering this picture, dark matter itself would have a lesser role in galaxies, while the RGGR role would be the dominant one.

Albeit the results here presented for NGC 4374 favor RGGR over MOND, a natural criticism would be that with RGGR  one more parameter is being fitted than with MOND. The comparison done here is worthwhile since: $i$) had we systematically found for RGGR a worse observational concordance than for MOND, that would  sign that RGGR had deep problems and that it should be dismissed; $ii$) in case MOND systematically deviates from the observational data (e.g., data from giant elliptical galaxies), it should be dismissed or modified. $iii$) MOND and RGGR seem to display  different tendencies for the stellar mass-to-light ratios, and thus developments on stellar population models may favor one over the other, in spite of any differences on the number of parameters.

\subsection{On the $\chi^2$ minimization} \label{Sec.chi2}
In order to find the best fit for each model, we proceed with a standard $\chi^2$ minimization. We numerically search for the global minimum of 

\be
	\chi^2(\Upsilon_*,  \Theta_1, \Theta_2) = \sum_{i =1}^N \( \frac{\sigma_{\mbox{\tiny model}}(\Upsilon_*,\Theta_1, \Theta_2, R_i)  - \sigma_{\mbox{\tiny obs}\, i}}{\zeta_i} \) ^2.
	\label{chi2}
\ee 
In the above, $N$ is the total number of observational VD points, $\Theta_1$  and $\Theta_2$ stand for additional parameters (e.g., $\beta$ and $\alpha \nu$), $\sigma_{\mbox{\tiny obs}\, i}$ is the VD observational value at radius $R_i$, while $\zeta_i$ is its corresponding uncertainty.

The procedure above only evaluates the shape of the VD curve, hence it may lead to small values of $\chi^2$ when the inferred mass-to-light ratios  are very far from the theoretical expectations. On the other hand, to simply fix the value of $\Upsilon_*$ according to a certain estimate seems (at least in many cases) unrealistic, since such estimates have significantly large error bars. Here, besides using the  standard minimization procedure as above, we also use a novel procedure in which the expected value of $\Upsilon_*$ is inserted as an additional data to be fitted, as follows,

\be
	\chi^2_\Upsilon (\Upsilon_*,  \Theta_1, \Theta_2) = \sum_{i =1}^N \( \frac{\sigma_{\mbox{\tiny model}}(\Upsilon_*,\Theta_1, \Theta_2, R_i)  - \sigma_{\mbox{\tiny obs}\, i}}{\zeta_i} \) ^2 + N \(\frac{\Upsilon_* - \bar \Upsilon_*}{\zeta_\Upsilon}\)^2,
	\label{chiUps}
\ee
where $\bar \Upsilon_*$ is the theoretically expected stellar mass-to-light ratio, $\zeta_\Upsilon$ is its error or dispersion, and the $N$  multiplying the last term imposes the same weigh for both the curve shape and the mass-to-light ratio.

One way to evaluate the goodness of the fit, which is probably the most straightforward,  is by the comparison of the $\chir$ values. In the tables \ref{N4374resultsTable} and \ref{N4494resultsTable} we list the values of the minimum  of $\chi^2$ and its corresponding $\chir$. The last is given by the minimum value of $\chi^2$ divided by the difference between $N$ and the number of parameters being fitted. When using the $\chi^2$ given by eq. (\ref{chiUps}), the number of observational data, $N$, is added by one.\footnote{Another reasonable option would be to add by $N$, but either choice would not change our conclusions.}

Besides determining the  parameters values for the best fit, we also find the confidence level curves for each model. The region inside the $n \; \sigma$'s  confidence level curve is such that $\chi^2 (p_1,...,p_k) \le \chi^2_{\mbox{\tiny min}} + \Delta \chi^2(n,k)$, where $\{p_i\}$ are the model's free parameters and the values of $\Delta \chi^2(n,k)$ can be found in standard references on statistics. This is important for testing the sensibility of each model and for disclosing correlations between its parameters.

As a final remark, likewise was done in Refs. \cite{2011MNRAS.411.2035N,2009MNRAS.393..329N}, the galaxy kinematics inside its 10'' radius is not considered at the fitting procedure. This is done since it is unlikely that the Jeans model employed is a reasonable approximation for the galaxy dynamics at a range so close to the galaxy center. In all of the VD plots the observational data is present from 5'', and the fitted curves are accordingly extended.

\subsection{Specific comments on the NGC 4374 and NGC 4494 fits}

\noindent
{\it NGC 4374}

\noindent
For Newtonian gravity and isotropic VD ($\beta = 0$), a mass model composed only of stars cannot fit the observational data of the galaxy NGC 4374, as it  it can be seen in Fig.\ref{N4374StarsPhotoFig} (see also Ref. \cite{2011MNRAS.411.2035N}). Besides being a poor fit considering the shape of the VD curve, it was achieved by using a stellar  mass-to-light ratio $\Upsilon_*$ which is considerably above the theoretical expectations. If the isotropy condition is dropped, the fit for the shape of the curve can be slightly improved, but  the anisotropy parameter goes toward the disfavored negative values $\beta \lesssim - 1$  and  $\Upsilon_*$ increases even more. This feature is common to other giant elliptical galaxies.

From Fig. \ref{N4374MONDFig} and Table \ref{N4374resultsTable}, it is clear that MOND fits better the NGC 4374 observational data than Newtonian gravity without dark matter. However, it is still a poor fit, in particular since: $i$) There is a significant tendency towards a lower VD curve at large radii (the region of PNe data), tendency which is strongly enhanced once the fits considers the expected $\Upsilon_*$ (the models iii and iv in Fig. \ref{N4374MONDFig}); $ii$) if the expected $\Upsilon_*$ is not used, the best fit is achieved for tangential anisotropy with  $\beta \le -1$. Also, besides these two points, MOND is incompatible with the Kroupa IMF expectations for this galaxy, since even for the model iv the derived $\Upsilon_*$ is (far) outside the $2 \sigma$ confidence level (Fig. \ref{N4374MONDFig}). Other  issues of MOND with the giant ellipticals can be found for instance in Ref.\cite{Gerhard:2000ck, Memola:2011vn}.

From Fig. \ref{N437RGGRPhotoFig} and Table \ref{N4374resultsTable}, it can be seen that RGGR fit to the data is a satisfactory one and outperforms MOND in all the points above.  It seems that the single issue that MOND does better than RGGR is on it's VD curve continuation towards the galactic center (i.e., the extension from 10'' (0.83 kpc) to 5'' (0.41 kpc)). Since the application of this Jeans modeling to a region so close to the galactic center is probably meaningless \cite{2011MNRAS.411.2035N}, that would constitute no true advantage to MOND.

\bigskip
\noindent
{\it NGC 4494}

\noindent
There is a wiggle at large radius in our analysis (Fig.\ref{N4494StarsPhotoFig}) that is not presented in Ref. \cite{2009MNRAS.393..329N}. This wiggle is innocuous to the fits, and it appears due to a different convention on the S\'ersic extension starting radius.

Similarly to other ordinary elliptical galaxies, mass models derived from Newtonian gravity and a stellar component  provide reasonable fits to the  observational VD data \cite{Romanowsky:2003qv, 2009MNRAS.393..329N}, se Table \ref{N4494resultsTable} and Fig. \ref{N4494StarsPhotoFig}. Actually, the fits with Newtonian gravity without dark matter are so good that they have lead to a conflict with CDM expectations \cite{Romanowsky:2003qv}. Such lack of dark matter-like effects was explained in the context of MOND shortly after \cite{Milgrom:2003ui}, since the typical internal accelerations of such galaxies are higher than MOND's $a_0$ acceleration. In the case of RGGR, considering that NGC 4374 is a much brighter elliptical, the expectation would be that NGC 4494's corresponding value of $\alpha$  would be significantly lower than that of the giant elliptical, while  probably (considering the disk galaxy fits) above $0.1/\nu$. Sharper expectations on the $\alpha$ of NGC 4494 can be drawn after similar galaxies have been analyzed in the RGGR framework. The resulting $\alpha$ value in Table \ref{N4494resultsTable} lies well inside that region. 

Likewise in the NGC 4374 case, the RGGR derived $\Upsilon_*$ in models i and ii (Fig.\ref{N4494RGGRPhoto3}) is naturally compatible with the Kroupa IMF expectations  (Table \ref{DLS}). Once such value enters as part of the data to be fitted, as in eq. (\ref{chiUps}) and models iii and iv, the parameters degeneracies  are significantly lowered, and sharper predictions are done.

\section{Conclusions}

In summary, in this paper  the RGGR \cite{Shapiro:2004ch, Rodrigues:2009vf} effective mass for stationary spherical systems was deduced, we presented general considerations on  elliptical galaxies within RGGR, and evaluated two specific ellipticals (an ordinary and a giant one), both with extended observed VD by the use of recent planetary nebulae (PNe) data \cite{2011MNRAS.411.2035N,2009MNRAS.393..329N}. Good agreement between the RGGR VD curve and the observations was found. Considering the galaxies here analyzed,  no strong tendency towards tangential anisotropy\footnote{Which  is usually disfavored, see also Ref.\cite{2012ApJ...748....2D}.} was  found for RGGR, while this behavior appeared in both Newtonian gravity without dark matter and MOND. The stellar mass-to-light ratios $\Upsilon_*$ found within RGGR are compatible with the expectations of the Kroupa IMF (or other similar IMF that lead to lower $\Upsilon_*$ than the Salpeter one), also in accordance with the findings of Ref. \cite{Rodrigues:2009vf}.

In the case of NGC 4374, the RGGR fit was clearly better than the MOND's one due to a number of features. The discrepancies between MOND and the observational data are significantly enhanced when a $\Upsilon_*$  compatible with the Kroupa IMF is assumed. 

In additional to the standard procedure of using $\Upsilon_*$ as a free parameter, and only evaluating  a posteriori whether  the fitting procedure yielded reasonable results, additional fits were generated in which the expected value of $\Upsilon_*$ was part of the data do be fitted. This procedure, in comparison with simply fixing the value of $\Upsilon_*$ a priori, is useful to avoid being unphysically precise on the value of $\Upsilon_*$.

The best way to compare different models is to use the same assumptions and procedures whenever possible. This methodology was  used here for Newtonian gravity, MOND and RGGR. In Refs. \cite{2009MNRAS.393..329N, 2011MNRAS.411.2035N} the NFW dark matter halo was analyzed for the same galaxies, but the higher order Jeans equations together with the kurtosis data were used as part of the data to be fitted.  In particular, the observational kurtosis can put constraints on the anisotropy parameter $\beta$, once the distribution function is assumed to satisfy $f(E,L) = f_0(E) L^{- 2 \beta}$ for constant $\beta$ (for further details, see the Appendix B of   Ref. \cite{2009MNRAS.393..329N} and references therein). In those references, the NFW halo fits were found to be compatible with the N-body simulations expectations and the Kroupa IMF, considering the effects of adiabatic contraction together with a crescent anisotropy along the radius (albeit with a low concentration value for NGC 4494). Comparing the shape of the VD curves,  the NFW halo could achieve about the same or a  better concordance with the observational data than RGGR (remember that NFW halos use one more free parameter than RGGR).  It is not easy to draw a straight comparison between the anisotropy parameters, but  RGGR has shown  a tendency towards isotropy while being only mildly dependent on the precise value of $\beta$ (since the 1 $\sigma$ uncertainties for RGGR are about or larger than 0.5). Had the kurtosis been evaluated as part of the data to be fitted, that would  probably insert a tendency towards higher radial anisotropy. Finally, considering the stellar mass-to-light ratios ($\Upsilon_*$), albeit both NFW and RGGR are compatible with the Kroupa IMF expectations,  RGGR shows a tendency in these galaxies towards lower $\Upsilon_*$ than NFW. And that may eventually constitute a physical test for RGGR, once the stellar populations models become more precise.

Another well known gravitational theory, $f(R)$ gravity, had its consequences for elliptical galaxies recently evaluated in Ref. \cite{2012ApJ...748...87N}. The sample of data that we use is similar to theirs. The VD curve  that they find is similar to the one derived for RGGR, while two clear differences are the $\Upsilon_*$ of NGC 4374 (RGGR has a lower value, more in the middle of the Kroupa IMF expectation) and the $\beta$ of NGC 4494 (since $f(R)$ shows  a significative tendency towards radial anisotropy).  It would be interesting to compare these models using a larger sample.

There is a number of natural  developments for RGGR, both from the theoretical side and the phenomenological one. In particular, within the context of galaxy kinematics, it is desirable to better constrain the $\alpha$ variation. This can be achieved from the analysis of a larger sample of galaxies. The $\alpha$  variation with the galaxy parameters is not currently well known,  but it should be stressed that RGGR without dark matter (or with a relatively small amount of it inside galaxies) imply the existence of  correlations between dark matter-like effects and baryonic matter. The existence of some correlations are well known for a long time, in particular here we briefly explored the correlations implied by the fundamental plane. It might result that no sufficiently strong correlation compatible with an $\alpha$ as a reasonable  function of $M_*, R_e, n$ to be found, hence galaxy kinematics alone (even if all the individual fits of galaxies are satisfactory) may refute, or significantly corroborate, this approach of RGGR without dark matter. Other developments  at cluster and cosmological scales and on lensing effects are being done.

\vspace{.2in}

\acknowledgments  
I am grateful to Ilya Shapiro for useful discussions on the renormalization group and for commenting on a preliminary version of this paper; to Nicola Napolitano for useful discussions on the analysis of elliptical galaxies and for providing data  on NGC 4374, NGC 4494; to Oliver Piattella and J\'ulio Fabris for important discussions on conformal transformations; and I also thank  CNPq and FAPES for partial financial support.

\appendix

\section{The $\Lambda$ variation inside galaxies} \label{appendixA}
The main purpose of this appendix is to show that, albeit $\Lambda$ is not constant inside galaxies in the RGGR context, its running is not sufficiently high to lead to any significative changes to the galaxy dynamics. The differential equation that governs the $\Lambda$ running, eq. (\ref{lambdalinha}), is below derived from the RGGR action and it is numerically evaluated for the galaxies NGC 4374 and NHC 4494, Fig. \ref{LambdaVariation}.

With the cosmological `constant' $\Lambda$, the action (\ref{rggraction}) reads,
\be
	S_{\mbox{\tiny RGGR}}[g] = \frac {c^3}{16 \pi }\int \frac {R - 2 \Lambda  } G \, \sqrt{-g} \,  d^4x.
	\label{rggractionfull}
\ee
The above, together with some  energy-momentum tensor $T_{\mu \nu}$, leads to the following field equations (see, e.g. \cite{Reuter:2003ca})
\be
	 G_{\mu \nu} +  \Lambda \,  g_{\mu \nu} + G \, \square G^{-1} g_{\mu \nu} - G \, \nabla_\mu \nabla_\nu G^{-1} = \frac{8 \pi G}{c^4} \, T_{\mu \nu} \, .
	\label{rggreq}
\ee

From the energy-momentum conservation ($T_{\mu \nu}^{\; \; \; ;\nu} = 0$), the Bianchi identities ($G_{\mu\nu}^{\;\;\; ; \nu} = 0$) and eq. (\ref{rggreq}), the divergent of the the terms with derivatives on $G$ in eq. (\ref{rggreq}) must vanish, hence
\be
	\nabla_\nu \( \frac{\Lambda}G \) = \frac 12  R \, \nabla_\nu G^{-1},
	\label{EMrggr}
\ee
since $\(  \square \nabla_\nu - \nabla_\nu \square \) G^{-1} = R_{\nu \kappa} \nabla^\kappa G^{-1}$. The above result is exact.

Equation (\ref{EMrggr})  also appeared in Refs. \cite{Koch:2010nn, Cai:2011kd}. In  \cite{Koch:2010nn} the authors  use this equation to motivate the RG energy scale  $\mu$  as a function of the Ricci scalar $R$.  The approach we follow here is different. Here  $\mu$ was chosen from physical considerations in the weak field limit (see the comments before and after eq. (\ref{murggr})), while the $G$ variation as a function of $\mu$ is given by eq. (\ref{gmu}).

In order to disclose the variation of $\Lambda$ as a function of $G$ and a matter content given by a dust component (i.e., $(T_{\mu}^\nu ) = \diag (- \rho \, c^2 \; \; 0 \; \; 0 \; \; 0)$), from the trace of eq.(\ref{rggreq}) we write,
\be
	R = \frac{8 \pi G}{c^2} \rho + 4 \Lambda + 3 G \square G^{-1}.
\ee
Replacing the above in eq. (\ref{EMrggr}), one finds
\be
	\nabla_\mu \Lambda= \( \frac{4 \pi G \rho}{c^2} + \Lambda + \frac 32 G \square G^{-1}\) G \nabla_\mu G^{-1}.
\ee
The above equation is here interpreted as the differential equation that defines the  $\Lambda$ variation. We do not have an analytical solution for $\Lambda$, but from a given $\rho$ and from the expression of $G$ as a function of the Newtonian potential (i..e, $G = G_0/(1 + 2 \nu \alpha \ln \Phi_N/\Phi_0)$, $\Lambda$ can be numerically determined. Namely,
\be
	\Lambda'= 2 \nu \alpha \frac{\Phi_N'}{\Phi_N} \[ \frac{4 \pi G \rho}{c^2} +  \Lambda + 3 \nu \alpha \(\frac{\nabla^2 \Phi_N}{\Phi_N} - \frac{\Phi_N'^2}{\Phi_N^2}\)\].
	\label{lambdalinha}
\ee

Contrary to $G$, the derivatives of $\Lambda$ do not appear in the field equations (\ref{rggreq}), hence $\Lambda$ can  be dynamically relevant only if its absolute value is sufficiently high. For comparison purposes, the $\Lambda$ value as estimated from $\Lambda$CDM cosmology is $\Lambda_{\Lambda\mbox{\tiny CDM}} = 1.3 \times 10^{-13} / \mbox{kpc}^2$, while $4 \pi G_0 \rho_*/c^2 \sim (10^{-5} - 10^{-11})/\mbox{kpc}^2$ inside elliptical galaxies (up to about 5 effective radius, with $\rho_*$ the stellar mass, no dark matter).

	To evaluate $\Lambda$ inside galaxies in the RGGR context the starting point is to fix the value of $\Lambda$ ``outside'' the galaxy, say at 25 effective radius ($R_e$). To generate the numerical solutions presented in Fig. \ref{LambdaVariation}, we use that $\Lambda (25 R_e ) = \Lambda_{\Lambda\mbox{\tiny CDM}}$. One can change the value of $\Lambda(25 R_e)$ by more than an order of magnitude without significative consequences to the numerical evaluations here presented. 
	
	In conclusion, within the RGGR framework it is indeed safe to neglect $\Lambda$ inside the galaxies NGC 4494 and NGC 4374 (and probably the same holds for any other galaxy).

\begin{figure}[thbp]
\begin{center}
	  \includegraphics[width=\textwidth]{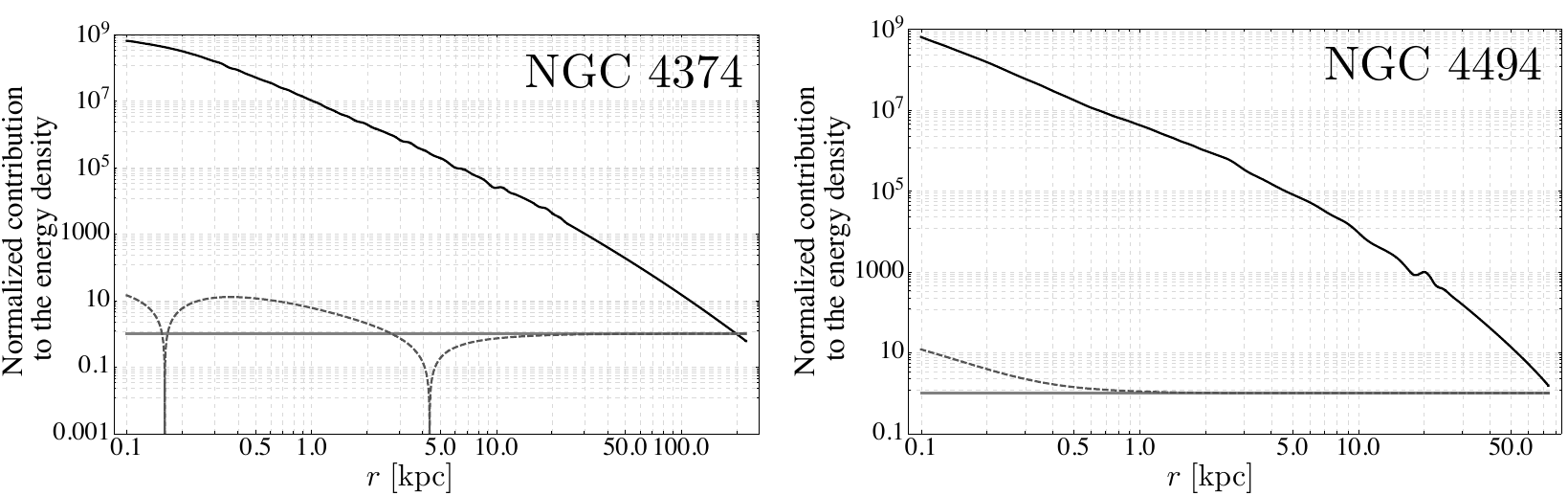}
\caption{A numerical comparison on the ratio of the absolute value of $\Lambda $ and $4 \pi G \rho_*/c^2$  to  $\Lambda_{0} = \Lambda(25 R_e)$ (see notes in text) ($\rho_*$ is the stellar mass density).  All parameters were fixed in accordance to one RGGR solution for  each galaxy (any of the solutions with different hypothesis on the baryonic content are compatible with the plots above).  The solid black curve corresponds to $4 \pi G\rho_*(r) /(c^2 \Lambda_0)$, the dashed gray curve  to $\Lambda(r)/\Lambda_0$, and the solid horizontal gray line has the constant value of 1. Inside the galaxy NGC 4374, $\Lambda$ becomes negative in a certain region and approaches the asymptotic value from below; while NGC 4494 approaches that value from above. This difference is due to the first galaxy being heavily driven by non-Newtonian dynamics (i.e., only very close to the center the right hand side of eq. (\ref{lambdalinha}) is negative). In both cases, the absolute value  of $\Lambda$, at any radius, is  too small to significantly change the average  internal galaxy dynamics.}
\label{LambdaVariation}
\end{center}
\end{figure}


\bibliographystyle{JHEP}

\bibliography{/Users/Davi/Desktop/Works/bibdavi2010}{}

\end{document}